\documentclass[%
        prd,
        aps,
        amsmath, amssymb,
       reprint,
       twocolumn,
       nofootinbib,
       superscriptaddress
    ]{revtex4-2} 
    
    \usepackage[utf8]{inputenc}
    \usepackage[T1]{fontenc}
\usepackage[english]{babel}
\usepackage{subcaption}
\usepackage{amsmath}
\usepackage{graphicx}
\usepackage{geometry}
\usepackage{url}
\usepackage{lipsum}
\usepackage{xcolor}
\usepackage{stmaryrd}
\usepackage{enumerate}

\usepackage{cleveref} 

\def\Msol{{M_\odot}}

\newcommand{\drond}[2][ ]{\frac{\partial #1}{\partial #2}}

\newcommand{\deriv}[2][ ]{\frac{\mathrm{d} #1}{\mathrm{d} #2}}
\newcommand{\derivdeux}[2][ ]{\frac{\mathrm{d}^2 #1}{\mathrm{d} #2^2}}

\renewcommand{\epsilon}{\varepsilon}

\begin{document}

\title{One- and two-argument equation of state parametrizations with continuous sound speed for neutron star simulations}

\author{Gaël Servignat}
\affiliation{Laboratoire Univers et Th\'eories, Observatoire de Paris, Universit\'e PSL, CNRS, Universit\'e Paris-Cit\'e, 92190 Meudon, France}
\author{Philip J. Davis}
\affiliation{Université de Caen Normandie, ENSICAEN, CNRS/IN2P3, LPC Caen UMR6534, F-14000 Caen, France}
\author{Jérôme Novak}
\author{Micaela Oertel}
\affiliation{Laboratoire Univers et Th\'eories, Observatoire de Paris, Universit\'e PSL, CNRS, Universit\'e Paris-Cit\'e, 92190 Meudon, France}
\author{Jos\'e A. Pons}
\affiliation{Departament de Física Aplicada, Universitat d'Alacant, Ap. Correus 99, E-03080 Alacant, Spain}

\begin{abstract}
  We describe two fitting schemes that aim to represent the
  high-density part of realistic equations of state for numerical
  simulations such as neutron star oscillations. The low-density part
  of the equation of state is represented by an arbitrary polytropic
  crust, and we propose a generic procedure to stitch any desired
  crust to the high-density fit, which is performed to ensure
  continuity of the internal energy, pressure and sound speed for
  barotropic equations of state that describe cold neutron stars in
  $\beta$-equilibrium. An extension of the fitting schemes to
  equations of state with an additional compositional argument is
  proposed. In particular we develop a formalism that ensures the
  existence of a $\beta$-equilibrium at low densities. An additional
  feature of this low-density model is that it can be, in principle,
  applied to any parametrization. The performance of the fits is
  checked on mass, radius and tidal deformability as well as on the
  dynamical radial oscillation frequencies. To that end, we use a
  pseudospectral single neutron star evolution code based on a
  non-conservative form of the hydrodynamical equations. A comparison
  to existing parametrizations is proposed, as far as possible, and to
  published radial frequency values in the literature. The static and
  dynamic quantities are well reproduced by the fitting schemes. Our
  results suggest that, even though the radius is very sensitive to
  the choice of the crust, this choice has little influence on the
  oscillation frequencies of a neutron star.
\end{abstract}

\maketitle

\section{Introduction}
When describing relativistic stellar structure and supernova or neutron star (NS) hydrodynamics, 
one needs to link the thermodynamic quantities using an equation of state (EoS). The choice of the EoS is crucial, 
mainly because the detailed physics at very high densities (i.e. where the strong nuclear interaction is the dominant one)
are poorly understood~\citep{oertel_equations_2017}. Polytropes are characterized by a high numerical precision on the computation
of thermodynamic variables due to their analytical nature, allowing them to be widely used in simulations~\citep{banyuls_numerical_1997,shibata_simulation_2000,kokkotas_radial_2001,font_three-dimensional_2002,cordero-carrion_improved_2009,hebert_general-relativistic_2018,rosswog_sphincs_bssn_2021}
but are however only very crude approximations for nuclear matter in the core or for the electron Fermi gas of the crust. 
Therefore, more realistic approaches of complex events such as a binary neutron star (BNS) merger need a detailed description of the strong nuclear interaction in the core.
These nuclear EoSs, the so-called realistic EoSs, often come as tables like those given in the \textsc{CompOSE} database\footnote{\protect\url{https://compose.obspm.fr}}~\citep{typel_compose_2015}.
Realistic EoSs are already used in BNS merger~\citep{shibata_merger_2005, hotokezaka_binary_2011, radice_binary_2018, zappa_binary_2023},
core-collapse~\citep{muller_new_2010, hempel_new_2012, just_core-collapse_2018, kuroda_core-collapse_2022},
proto-neutron star cooling~\citep{pascal_proto-neutron_2022, beznogov_standard_2023} and General Relativistic magneto-hydrodynamic dynamo codes~\citep{kiuchi_large_2023}.

However, these tables may {induce numerical artifacts, degrading
  the overall accuracy of the simulation}. First, due to the
non-analytical nature of nuclear models at high densities, the
precision on the thermodynamic quantities is often far from the
typical machine accuracy used in computers. {This may have an
  impact on the computation of derivatives of the EoS, like the sound
  speed, or those that are necessary for interpolation. Computing the
  sound speed with finite-differences schemes can present non-physical
  spikes that may lead to code failure.} Secondly, increasing the
complexity of physical hypotheses leads to EoSs with two
or three arguments, namely by considering composition and
temperature effects, meaning that even with a reasonable amount of
discretization points for every argument (of the order of a hundred),
the tables may contain millions of entries, and the associated files
would be very large. Some EoSs even depend upon more
than three arguments if more particles, such as muons, are
included. Moreover, computing the thermodynamic quantities at non
tabulated values has to be done with interpolation, which is in itself
not a trivial problem. For example,
Swesty~\citep{swesty_thermodynamically_1996} has proposed a way of
interpolating tables in a thermodynamically consistent manner using
Hermite cubic and quintic splines. However, this technique can suffer
from spurious oscillations of higher-order polynomials, especially at
low densities where the derivatives can be the noisiest. Representing
a table analytically is a way to avoid this problem. Several
analytical representations of tabulated EoSs already
exist, such as piecewise polytropes~\citep{read_constraints_2009} that
were later improved~\citep{suleiman_polytropic_2022} and generalized
(hereafter referred to as GPP)~\citep{oboyle_parametrized_2020},
spectral representations~\citep{lindblom_spectral_2010,
  lindblom_causal_2018, lindblom_improved_2022}, as well as the more
recent parametrizations of the sound speed~\citep{greif_equation_2018}
and of the enthalpy~\citep{legred_simulating_2023}.

These representations are all dedicated to cold, $\beta$-equilibrated
EoSs which only depend on one thermodynamic argument, typically
chosen as the baryon number density or the (pseudo)-enthalpy. Although
not a parametrization,~\citep{raithel_finite_2019} gives an analytical
extension of cold EoSs to arbitrary proton fraction outside of
$\beta$-equilibrium and to finite temperatures. We here propose an
analytical representation dubbed \textit{pseudo-polytrope}, both for one-
and two-argument EoSs{, with corresponding analytic models for
  low-density parts (crust), ensuring $\beta$-equilibrium in the
  two-argument case}. To benchmark our approach for one-argument
EoSs, we will compare the performances of the pseudo-polytrope with a
second type based on the analytical approach of Potekhin et
al.~\citep{potekhin_analytical_2013} and Pearson et
al.~\citep{pearson_analytical_2018} that was designed for a very
precise representation of Brussels-Montreal unified equations of
state. In both cases, the fits are performed in a high-density
interval corresponding to homogeneous matter in the core, and the
low-density part is attatched in a thermodynamically consistent
manner. The fits are then tested on static quantities such as the
mass, radius and tidal deformability, and on dynamic quantities,
namely the radial frequencies of spherically symmetric stars, as
computed with the code that has been presented
in~\citep{servignat_new_2023}. As such, the code only takes
one-argument EoSs, but to take electrons into account
only the evolution equation for the electron fraction has been added
in the code when using two-argument EoSs, making the generalization
straightforward.

We can note here an alternative approach to include out of
$\beta$-equilibrium effects in hydrodynamics simulations, as done in
recent works on the so-called bulk viscosity to provide effective ways
to describe such fluids out of the weak $\beta$-equilibrium in General
Relativity~\citep{celora_formulating_2022, camelio_simulating_I_2023,
  camelio_simulating_II_2023}. Nevertheless, in this work we follow
the transport approach in which the potential absence of weak
equilibrium is considered through a separate conservation equation for
the electron fraction that contains source terms to take neutrino
production into account. The source terms are computed thanks to the
description of neutrino emission rates found
in~\citep{haensel_non-equilibrium_1992}.

To test our fits we choose three different EoS models,
covering different techniques and a relatively large range of neutron
star global properties. One model is based on relativistic density
functional theory (DFT), one is based on a Skyrme (non-relativistic)
density functional and one on an empirical extension of a variational
microscopic model. All of them are reasonably compatible with existing
constraints from nuclear experiments, theory, and astrophysics. To be
specific, for the two-argument fits we consider the lowest
temperature entry of the general purpose EoS models~:
(i)~RG(SLy4)~\citep{gulminelli_unified_2015,raduta_nuclear_2019}, a {nucleonic}
non-relativistic DFT one; (ii)~the {nucleonic} relativistic DFT one
HS(DD2)~\citep{hempel_statistical_2010,typel_composition_2010}; as
well as (iii)~the SRO(APR)
model~\citep{schneider_open-source_2017,schneider_akmal_2019}. The
latter is based on the APR EoS~\citep{akmal_equation_1998}, which
itself is partly adjusted to the variational calculation 
of~\citep{akmal_spin_1997}{, and contains a phase transition to a pion condensate at high densities}. For the corresponding one-argument fits, we use
the zero temperature version of the {nucleonic} EoS
RG(SLy4)~\citep{gulminelli_unified_2015, chabanat_skyrme_1998,
  danielewicz_symmetry_2009} and the APR(APR)
EoS~(Chapter 5.12 of~\citep{haensel_neutron_2007}, based 
on~\citep{akmal_equation_1998}). {The latter uses a mixed phase to describe the transition to the pion condensate which therefore smoothens the EoS.} The fitting coefficients for the {nucleonic} GPPVA(DD2)
EoS~\citep{typel_composition_2010,
  grill_equation_2014,pearson_unified_2018} are also provided. The
data, as well as references and details for all EoSs are publicly
available in tabulated form from the \textsc{CompOSE}
database~\citep{typel_compose_2015}; the naming convention is the same as in
this database, too.

Throughout the whole paper we use geometrized units where $c=1$ and
$G=1$, except in Sec.~\ref{subsec:fitspearson}.

The paper is organised as follows: sections~\ref{sec:oneparameos} and~\ref{sec:twoparameos} are dedicated to the description of procedures
to represent one- and two-argument EoSs,
respectively. Results are presented in section~\ref{sec:results} with
a comparison to static quantities (maximum mass, tidal deformability)
and dynamic quantities (radial oscillation frequencies) both against
published values and between the two fitting schemes. In particular a
study of the influence of the choice of the crust is shown, as well as
a comparison between frequencies of one- and two-argument
EoSs. Conclusions are drawn in section~\ref{sec:conclusion}.

\section{Representation of one-argument equations of state}\label{sec:oneparameos}

\begin{figure*}
    \centering
    \includegraphics[width=\textwidth]{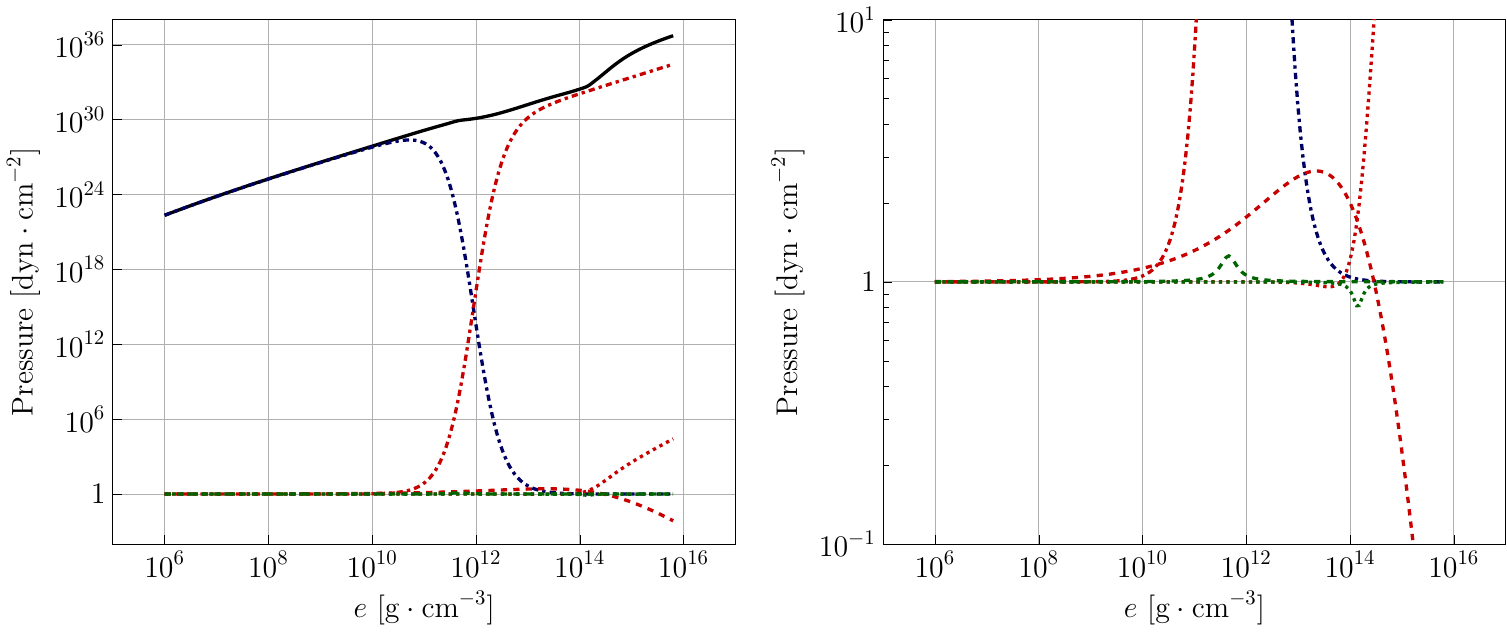}
    \caption{Contribution of each of the terms in Eq.~(\ref{eqn:pearsonfit}) to the total fit (solid, black curve) for the BSk24 EoS, using the fit coefficients presented in \citep{pearson_analytical_2018}. The right panel is a zoom on the lower part of the left panel. Term 1: blue, dot-dashed; term 2: red, dot-dashed; term 3: red, dotted; term 4: red, dashed; term 5: green, dashed, term 6: green, dotted.
    }
    \label{fig:bsk24_fit}
  \end{figure*}
  
This section is devoted to the presentation of the fitting schemes to
parametrize barotropic EoSs that describe cold neutron stars in
$\beta$-equilibrium. Two schemes are presented and for each one, the
global strategy is the following: the fitting scheme is applied to the
core of the NS, i.e. for densities above some threshold
$n_{\mathrm{lim},2}$; for densities below $n_{\mathrm{lim},1}$ a
polytrope is considered for the crust. In between, a GPP is used to
get a continuous matching of thermodynamic quantities: energy density $e$,
pressure $p$ and sound speed $c_s$, both at $n_{\mathrm{lim},1}$ and
$n_{\mathrm{lim},2}$. 

\subsection{Pseudo-polytropes}\label{subsec:pseudopoly}
We start with a fitting functional that we call \textit{pseudo-polytropes}. 
If $\epsilon = e/m_Bn_B - 1$ is the rescaled internal energy per particle (excluding rest-mass energy), with $n_B$ the baryon number density, $m_B$ the baryon mass and $e$ the total energy density, then the basic functional is:
\begin{equation}
	\label{eq:pseudopolytrope}
	\epsilon(n_B) = g(n_B)n_B^\alpha,
\end{equation}
where $g$ is an arbitrary function. The name pseudo-polytrope is justified by the following: the choice of $g=\kappa/m_B\alpha$ where $\kappa$ is a constant and denoting $\gamma = \alpha + 1$ yields a polytrope: 
\begin{equation}
	p=n_B^2\left(\drond[\epsilon]{n_B}\right)_{T,Y_e}= n_B^2\deriv[\epsilon]{n_B} = \kappa n_B^\gamma.
\end{equation}
where $p$ is the pressure. In order to account for an arbitrary crust, following the approach by~\citep{oboyle_parametrized_2020}, we add two parameters $L$ and $d$ such that the final functional is:
\begin{equation}
	\epsilon(n_B) = g(n_B)n_B^\alpha + d - \frac{L}{n_B}.
\end{equation}
The fitting coefficients will be $\alpha$ and the coefficients $\{\bar{a}_{i}\}_{i\in\llbracket0,n\rrbracket}$ of $g$, that we choose to be a polynomial:
\begin{equation}\label{eq:coefsg}
	g(x=\ln(n_B\,[\mathrm{fm}^{-3}])) = \sum\limits_{i=0}^{n}\bar{a}_ix^i~.
\end{equation}
Their values will be adjusted by fitting the functional and its
derivatives, whereas most parametrizations use only a single thermodynamic quantity to perform the fits.

From now on, $x$ denotes the natural logarithm of the
density $n_B$ expressed in fm$^{-3}$. In order to perform a
$\mathcal{C}^2$ stitching of the fit and an arbitrary crust, i.e. with
continuous sound speed, we add a single intermediate density interval that
is described by GPP formalism:
\begin{align}
  e_\mathrm{GPP}(n_B) & = \frac{K_\mathrm{GPP}}{\Gamma_\mathrm{GPP}-1}n_B^{\Gamma_\mathrm{GPP}} + (1+d_\mathrm{GPP})n_B \nonumber \\
                      & - L_\mathrm{GPP}, \\
  p_\mathrm{GPP}(n_B) & = K_\mathrm{GPP}n_B^{\Gamma_\mathrm{GPP}} + L_\mathrm{GPP}.
\end{align}
Then, for a given crust defined by pressure and energy density
profiles $p_c(n_B),\,e_c(n_B)$, there are three junction conditions at two densities $n_{\mathrm{lim},i}$, 
$i=1,2$, with $n_{\mathrm{lim},1}<n_{\mathrm{lim},2}$ that allow to determine the six parameters $K_\mathrm{GPP},\,\Gamma_\mathrm{GPP},\,d_\mathrm{GPP},\,L_\mathrm{GPP},\,d,\,L$:
\begin{widetext}
  \begin{align}
    K_\mathrm{GPP}\Gamma_\mathrm{GPP}n_{\mathrm{lim},1}^{\Gamma_\mathrm{GPP}-1}   & = \deriv[p_c]{n_B}(n_{\mathrm{lim},1}), \\
    K_\mathrm{GPP}\Gamma_\mathrm{GPP}n_{\mathrm{lim},2}^{\Gamma_\mathrm{GPP}}     & = n_{\mathrm{lim},2}\left(\deriv[(g(x)e^{\alpha x})]{x}(n_{\mathrm{lim},2}) + \derivdeux[(g(x)e^{\alpha x})]{x}(n_{\mathrm{lim},2})\right), \\
    K_\mathrm{GPP}n_{\mathrm{lim},1}^{\Gamma_\mathrm{GPP}} + L_\mathrm{GPP} & = p_c(n_{\mathrm{lim},1}), \\
    K_\mathrm{GPP}n_{\mathrm{lim},2}^{\Gamma_\mathrm{GPP}} + L_\mathrm{GPP} & = n_{\mathrm{lim},2}\deriv[(g(x)e^{\alpha x})]{x}(n_{\mathrm{lim},2}) + L, \\
    \frac{K_\mathrm{GPP}}{\Gamma_\mathrm{GPP}-1}n_{\mathrm{lim},1}^{\Gamma_\mathrm{GPP}} + (1+d_\mathrm{GPP})n_{\mathrm{lim},1} - L_\mathrm{GPP} & = e_c(n_{\mathrm{lim},1}),\\
    \frac{K_\mathrm{GPP}}{\Gamma_\mathrm{GPP}-1}n_{\mathrm{lim},2}^{\Gamma_\mathrm{GPP}-1} + d_\mathrm{GPP} - \frac{L_\mathrm{GPP}}{n_{\mathrm{lim},2}} 
                                                                                  & = g(x_{\mathrm{lim},2})e^{\alpha x_{\mathrm{\lim},2}} + d - \frac{L}{n_{\mathrm{lim},2}}.
\end{align}
\end{widetext}
They correspond, by groups of two, to the continuity conditions of the internal energy and its first two derivatives. 
The first two equations are independent of $d_\mathrm{GPP},\,L_\mathrm{GPP},\,d,\,L$ and are solved with the non-linear vector function root-finder \texttt{root} of the \texttt{scipy.optimize} Python library~\citep{virtanen_scipy_2020} to determine $K_\mathrm{GPP}$ and $L_\mathrm{GPP}$.
Only those first two equations are coupled, while the subsequent four equations can be solved successively to directly determine $L_\mathrm{GPP},\,L,\,d_\mathrm{GPP},\,d$ in this order.
The fit is performed with a minimization of the following cost function:
\begin{widetext}
\begin{equation}
	\label{eq:costfunction}
	E(\{\bar{a}_j\},\alpha,d,L) = {\frac{1}{x_\mathrm{max}-x_\mathrm{min}}\sum\limits_{X}\sum\limits_{i}\left(\frac{X_{\mathrm{tab},i} - X_\mathrm{fit}(x_i, \{\bar{a}_j\},\alpha,d,L)}{X_{\mathrm{tab},i}}\right)^2\Delta{x}_i},
\end{equation}
\end{widetext}
where $x_\mathrm{max,min}$ are the limiting densities of the chosen fitting interval, $\Delta{x}_i$ is the log-density step of the table, i.e. if the densities of the table are discretized as $\{x_i,\,i\in[0,N]\}$ where $x_0=x_\mathrm{min}\leq x_1\leq \dots\leq x_{N-1}\leq x_N = x_\mathrm{max}$, then $\Delta{x}_i = x_{i+1} - x_i$ when $i=0,\dots,N-1$, and $\Delta{x}_N = \Delta{x}_{N-1}$ and the variable $X$ spans the chosen quantities to fit on, making the procedure a joint fit. For the one-argument fits the fit is performed on $\epsilon/n_B$, $\ln(p/n_B)$ and $\Gamma_1$ (defined by Eq.~\ref{e:def_Gamma1}), which are chosen to represent the internal energy $\epsilon$ and its first two derivatives, rescaled so that they are all of close order of magnitude.

We start by producing a set of fit coefficients with $d=L=0$,
thanks to a linear least-square fit of $\epsilon$ with respect to $x$ performed with the \texttt{curve\_fit} routine of \texttt{scipy.optimize}~\citep{virtanen_scipy_2020}.
This educated guess is then used as the starting point of the minimization of Eq.~\eqref{eq:costfunction}. At every step of the procedure, $K_\mathrm{GPP},\,\Gamma_\mathrm{GPP},\,d_\mathrm{GPP},\,L_\mathrm{GPP},\,d,\,L$ can be computed from the current value of the coefficients.
In practice, the crust is taken to be a polytrope for which the parameters $(\kappa,\gamma)$ are chosen freely:
\begin{equation}
	p_c(n_B) = \kappa n_B^\gamma.
\end{equation}
Once $\gamma$ is chosen, a fine-tuning of $\kappa$ is done by hand to ensure that the maximal mass of the EoS is recovered. The results of the fit are reported in Table~\ref{tab:coefpseudopolytrope} of App.~\ref{s:fit_coefs} for one-argument fits of RG(SLy4)~\citep{gulminelli_unified_2015, chabanat_skyrme_1998, danielewicz_symmetry_2009}, APR(APR)~\citep{haensel_neutron_2007, akmal_equation_1998} and GPPVA(DD2)~\citep{typel_composition_2010, grill_equation_2014,pearson_unified_2018}.

\begin{figure*}
  \begin{minipage}[t]{0.95\textwidth}
  \begin{subfigure}[b]{.47\textwidth}
    \centering
    \includegraphics[width=\columnwidth]{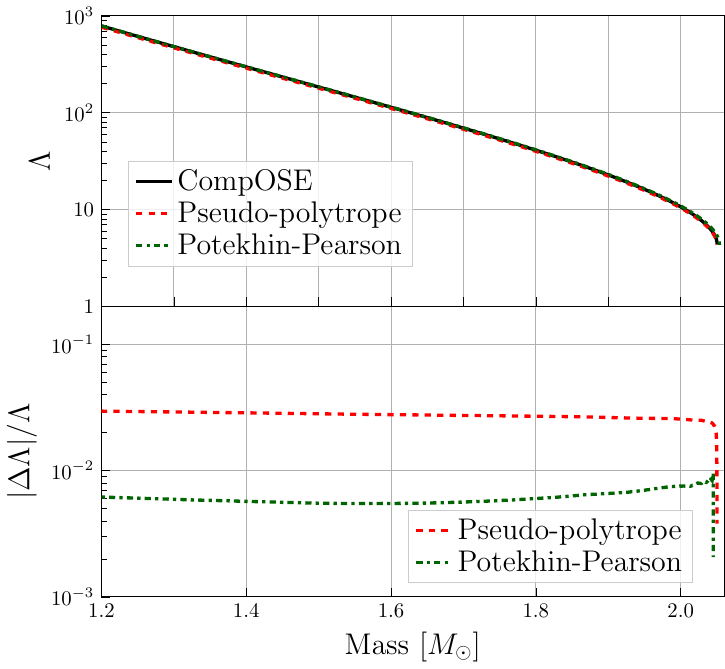}
    \caption{RG(SLy4)~\citep{gulminelli_unified_2015, chabanat_skyrme_1998, danielewicz_symmetry_2009}.}
    \label{fig:lmdiagSLy4}
  \end{subfigure}
  \hfill
  \begin{subfigure}[b]{.47\textwidth}
    \centering
    \includegraphics[width=\columnwidth]{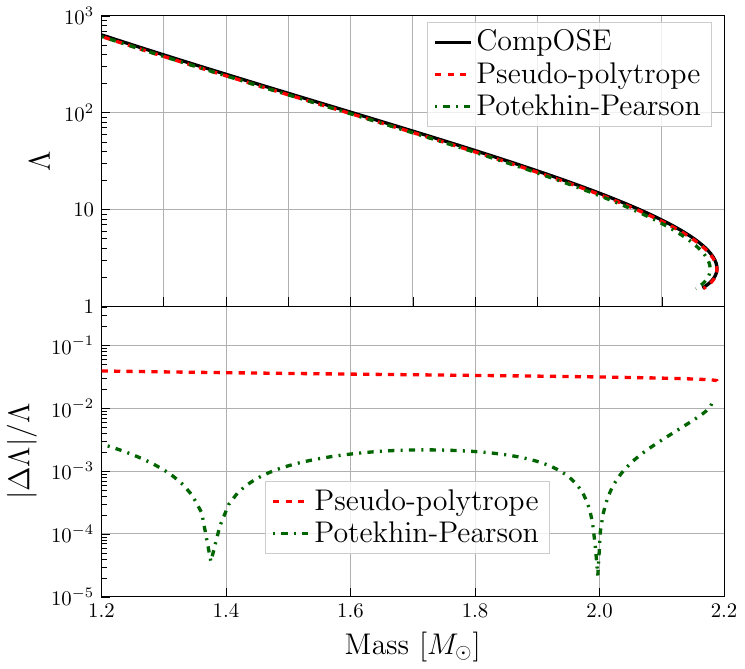}
    \caption{APR(APR)~\citep{haensel_neutron_2007, akmal_equation_1998}.}
    \label{fig:lmdiagAPR}
  \end{subfigure}
  \caption{Mass-deformability diagrams of the two cold EoSs. The bottom panel is the relative difference of the fits with respect to the original \textsc{CompOSE} EoS. {The relative differences have been computed at constant fraction of the maximum mass: $\Lambda_\mathrm{EoS}(M_1)$ was compared to $\Lambda_\mathrm{fit}(M_2)$ where $M_2$ is defined by ${M_2}/{M_{\mathrm{fit},\mathrm{max}}} = {M_1}/{M_{\mathrm{EoS},\mathrm{max}}}$}.}
  \label{fig:lmdiag1param}
  \end{minipage}
\end{figure*}

\begin{figure*}
  \begin{minipage}[t]{0.95\textwidth}
  \begin{subfigure}[b]{.29\textwidth}
    \centering
    \includegraphics[width=\columnwidth]{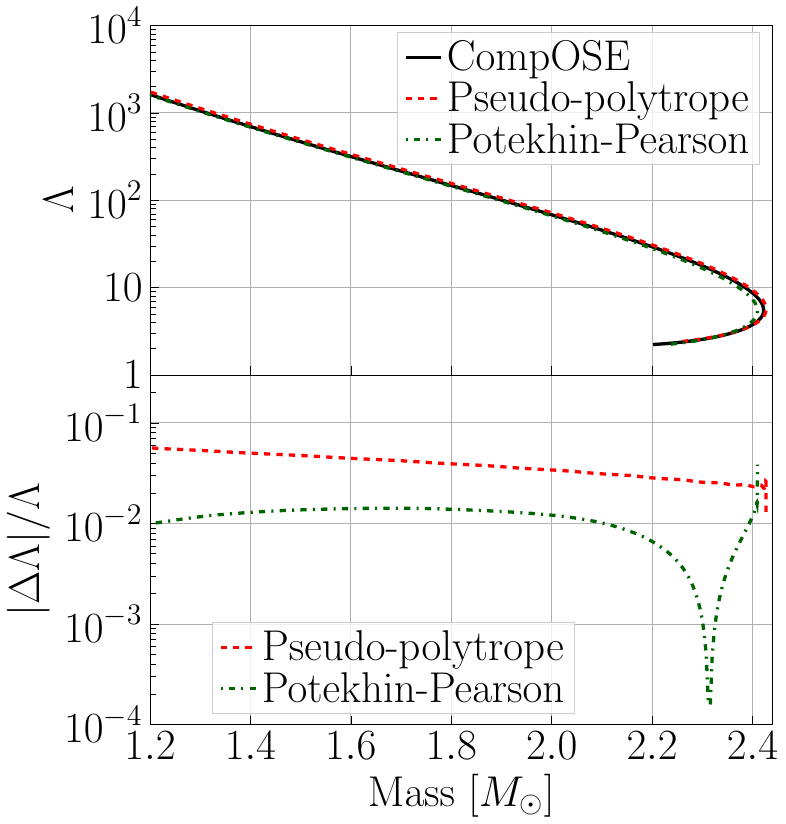}
    \caption{HS(DD2)~\citep{hempel_statistical_2010,typel_composition_2010}.}
    \label{fig:lmdiagDD2}
  \end{subfigure}
  \hfill
  \begin{subfigure}[b]{.30\textwidth}
    \centering
    \includegraphics[width=\columnwidth]{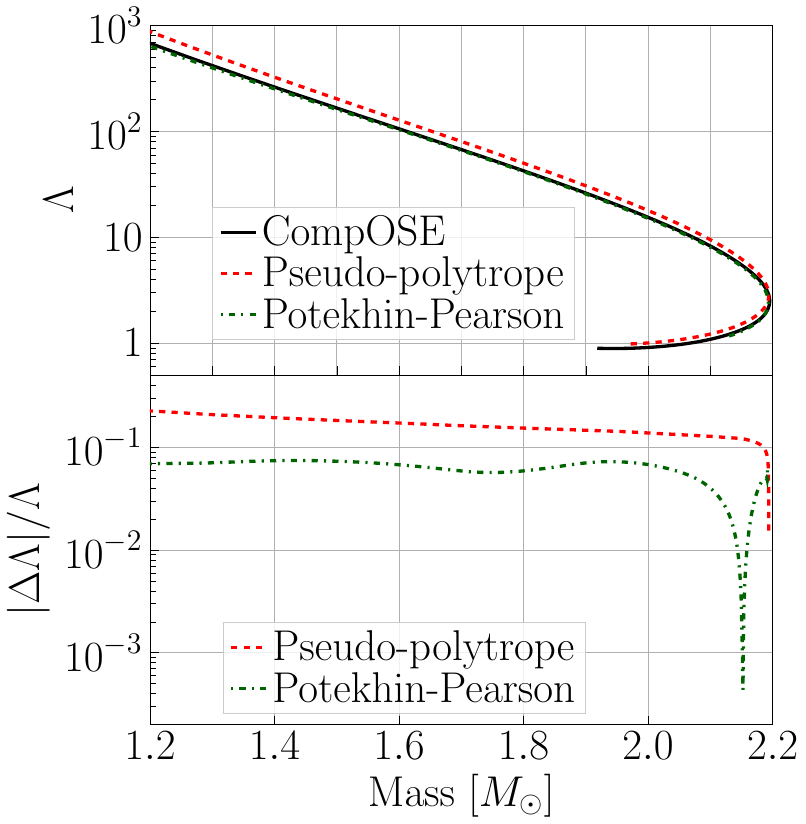}
    \caption{SRO(APR)~\citep{schneider_open-source_2017,schneider_akmal_2019}.}
    \label{fig:lmdiagSROAPR}
  \end{subfigure}
  \hfill
  \begin{subfigure}[b]{.29\textwidth}
    \centering
    \includegraphics[width=\columnwidth]{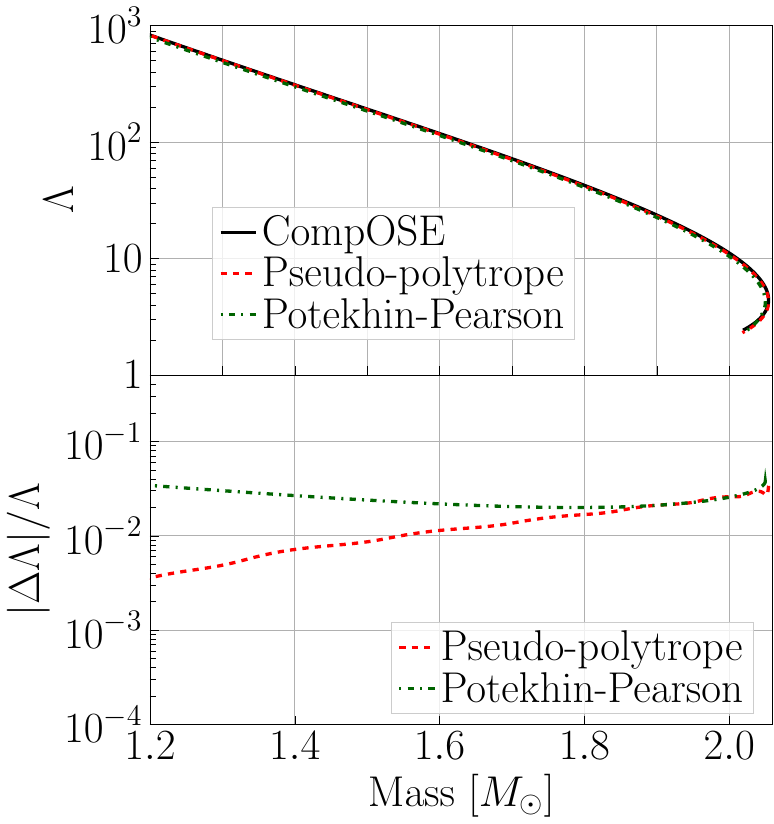}
    \caption{RG(SLy4)~\citep{gulminelli_unified_2015, raduta_nuclear_2019}.}
    \label{fig:lmdiagRGSLy4}
  \end{subfigure}
  \caption{Mass-deformability diagrams of the $\beta$-equilibrated versions of the three general purpose EoSs. The bottom panel is the relative difference of the fits with respect to the original \textsc{CompOSE} EoS. The relative differences have been computed at constant fraction of the maximum mass: {$\Lambda_\mathrm{EoS}(M_1)$ was compared to $\Lambda_\mathrm{fit}(M_2)$ where $M_2$ is defined by ${M_2}/{M_{\mathrm{fit},\mathrm{max}}} = {M_1}/{M_{\mathrm{EoS},\mathrm{max}}}$}.}
  \label{fig:lmdiag2param}
  \end{minipage}
  \end{figure*}

  \subsection{The Potekhin-Pearson fitting scheme}\label{subsec:fitspearson}

  \citep{potekhin_analytical_2013} and~\citep{pearson_analytical_2018} presented an analytical representation between the log of the pressure, $\mathrm{log}_{10}p\equiv{\zeta}$, and the log of the total energy density, $\mathrm{log}_{10}\left(\hat{e}\,[\mathrm{g}\cdot\mathrm{cm}^{-3}]\right)\equiv\xi$ according to
  \begin{equation}
    \begin{aligned}
      \zeta & = K+\frac{a_{1}+a_{2}\xi+a_{3}\xi^{3}}{1+a_{4}\xi}f_{0}(a_5(\xi-a_6)) \\
      & + (a_7+a_{8}\xi)f_{0}(a_{9}(a_{6}-\xi)) \\
      & + (a_{10}+a_{11}\xi)f_{0}(a_{12}(a_{13}-\xi)) \\
      & + (a_{14}+a_{15}\xi)f_{0}(a_{16}(a_{17}-\xi)) \\
      & + \frac{a_{18}}{1+[a_{20}(\xi-a_{19})]^{2}} \\
      & + \frac{a_{21}}{1+[a_{23}(\xi-a_{22})]^{2}},\\
    \end{aligned}
    \label{eqn:pearsonfit}
  \end{equation}
  where $f_{0}(x)=\left\{\exp(x)+1\right\}^{-1}$, $a_{i}$ are the fitting coefficients and $\hat{e} = e / c^2$ is in g cm$^{-3}$. Setting $K=0$ gives the pressure in units of dyn cm$^{-2}$, while $K=-33.2047$ gives the pressure in MeV fm$^{-3}$.
  
  \citep{pearson_analytical_2018} showed that Eq.~\eqref{eqn:pearsonfit} could calculate the pressure with typical errors of about one per cent for the Brussels-Montreal Skyrme (BSk) functionals, for densities in the range $6\lesssim{\xi}\lesssim{16}$. Indeed, each of the terms in Eq.~\eqref{eqn:pearsonfit} address a specific region of the NS (see Fig.~\ref{fig:bsk24_fit}); the first term is associated with the outer crust while the second, third and fourth terms describe the inner crust and the core regions. The fifth and sixth terms describe the neutron drip and the core-crust boundary, respectively.
  
  Despite the fact that Eq.~\eqref{eqn:pearsonfit} can compute the pressure over the whole NS domain, we neglect the terms associated with the crust since our aim is to model the high-density part of the EoSs only.
  We therefore apply the analytical fit
  \begin{equation}
    \begin{aligned}
      \zeta & = K + (a_2+a_{3}\xi)f_{0}(a_{4}(a_{1}-\xi))\\
      & + (a_{5}+a_{6}\xi)f_{0}(a_{7}(a_{8}-\xi)) \\
      & + (a_{9}+a_{10}\xi)f_{0}(a_{11}(a_{12}-\xi)) \\
      & + \frac{a_{13}}{1+[a_{15}(\xi-a_{14})]^{2}}
    \end{aligned}
    \label{eqn:pearsonfit_core}
  \end{equation}
  to the CompOSE tabulated values of pressure and total energy densities where $n_{\mathrm{B}}\geq{0.05}$ fm$^{-3}$. This lower limit in $n_{\mathrm{B}}$ for the fitting window was chosen by hand to ensure good quality fits.
  In accordance with the GPP approach the pressure is then computed with
  \begin{align}
    p(e) & = 10^{\zeta(\xi)} + L.
  \end{align}
  
  The crust model is then added as described in Sec.~\ref{subsec:pseudopoly}. The equations describing the continuity of the pressure gradient, the pressure and the energy density at the boundary $n_{\mathrm{lim,2}}$ for the Potekhin-Pearson scheme are, respectively
  \begin{widetext}
    \begin{align}
      K_{\mathrm{GPP}}\Gamma_{\mathrm{GPP}}n_{\mathrm{lim,2}}^{\Gamma_{\mathrm{GPP}}-1} & =\frac{\mathrm{d}p(n_{\mathrm{lim,2}})}{\mathrm{d}n_{\mathrm{B}}}=\frac{p(e_{\mathrm{lim,2}})+e_{\mathrm{lim,2}}}{n_{\mathrm{lim,2}}}\frac{p(e_{\mathrm{lim,2}})}{e_{\mathrm{lim,2}}}\frac{\mathrm{d}\zeta}{\mathrm{d}\xi},\\
      K_{\mathrm{GPP}}n_{\mathrm{lim,2}}^{\Gamma_{\mathrm{GPP}}}+L_{\mathrm{GPP}} & =p(e_{\mathrm{lim,2}})+L,\\
      \frac{K_\mathrm{GPP}}{\Gamma_\mathrm{GPP}-1}n_{\mathrm{lim},2}^{\Gamma_\mathrm{GPP}} + (1+d_\mathrm{GPP})n_{\mathrm{lim},2} - L_\mathrm{GPP} & = e_{\mathrm{lim,2}} + (1+d)n_{\mathrm{lim,2}}-L.
    \end{align}
  \end{widetext}
  Here, $\mathrm{d}\zeta/\mathrm{d}\xi$ is calculated from Eq.~\eqref{eqn:pearsonfit_core} using the \texttt{sympy} library for symbolic computation, $e_{\mathrm{lim,2}}$ is the {total} energy density at $n_{\mathrm{lim,2}}$ and $p(e_{\mathrm{lim,2}})$ is the pressure at this location, again calculated with Eq.~\eqref{eqn:pearsonfit_core}. The value of $e$ for a given $n_{\mathrm{B}}$ is calculated by inverting
  \begin{equation}
    \mathrm{ln}\left(\frac{n_{\mathrm{B}}}{n_{0}}\right)=\int_{e_{0}}^{e}\frac{\mathrm{d}e^{\prime}}{p(e^{\prime})+e^{\prime}},
    \label{eqn:solve_nB}
  \end{equation}
  {(see Eq. (3) of~\citep{haensel_analytical_2004})} where $e_{0}$ is the first data point in the fitting interval for the {total} energy density, and $n_{0}$ is the initial baryon density
  which, in turn, is calculated from the definition of the enthalpy
  \begin{equation}
    n_{0}=\frac{e_{0}+p(e_{0})}{h_{0}},
    \label{eqn:h0}
  \end{equation}
  where $h_{0}$ is the first data point in the fitting interval for the enthalpy per baryon, and is calculated from the tables provided for a given EoS using the \textsc{CompOSE} software, and $p(e_{0})$ is calculated from the fit.
    We use the values of $\kappa$, $\gamma$ and
  $n_{\mathrm{lim},1}$ for the crust model presented in
  Sec.~\ref{subsec:pseudopoly}
  (cf. Table~{\ref{tab:parampseudopolytrope}} of App.~\ref{s:fit_coefs}). However, we fine tuned
  the values of $n_{\mathrm{lim},2}$ to ensure good quality fits. We
  use 0.04 fm$^{-3}$, 0.03 fm$^{-3}$ {and 0.04 fm$^{-3}$} for the barotropic
  RG(SLy4)~\citep{gulminelli_unified_2015, chabanat_skyrme_1998,
    danielewicz_symmetry_2009}, APR(APR)~\citep{haensel_neutron_2007, akmal_equation_1998} and {GPPVA(DD2)~\citep{typel_composition_2010,grill_equation_2014,pearson_unified_2018}} respectively. For the two-argument EoSs
  RG(SLy4)~\citep{gulminelli_unified_2015, raduta_nuclear_2019},
  SRO(APR)~\citep{schneider_open-source_2017,schneider_akmal_2019} and
  HS(DD2)~\citep{hempel_statistical_2010,typel_composition_2010} we
  use 0.07 fm$^{-3}$, {0.05} fm$^{-3}$ and 0.04 fm$^{-3}$,
  respectively. The results of the fit for the barotropic
  RG(SLy4)~\citep{gulminelli_unified_2015, chabanat_skyrme_1998,
    danielewicz_symmetry_2009} and
  APR(APR)~\citep{haensel_neutron_2007, akmal_equation_1998} are given
  in Table~\ref{tab:coefpearson} of App.~\ref{s:fit_coefs}.
  
  \begin{figure}
    \centering
    \includegraphics[width = \columnwidth]{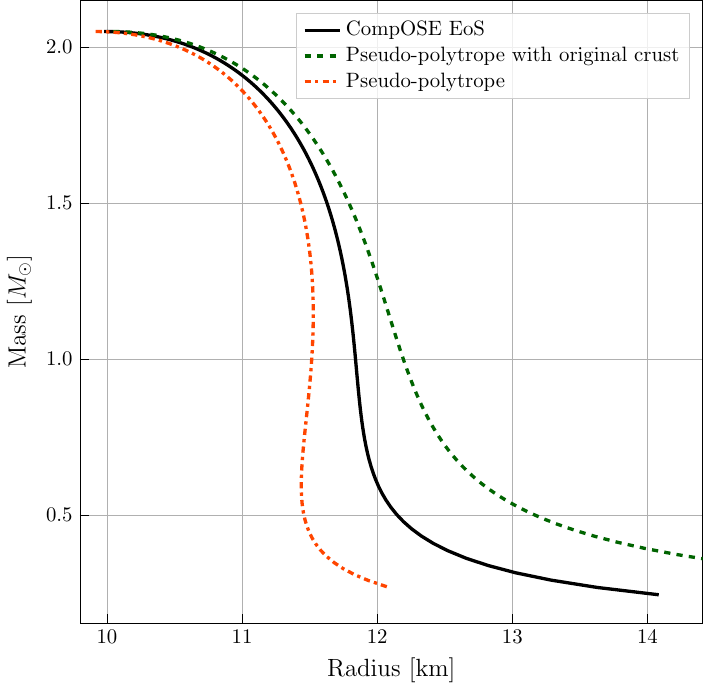}
    \caption{Influence of the crust on the radius. The solid line is the
      \textsc{CompOSE} EoS. The dashed line is a pseudo-polytropic fit, and the dash-dotted line is the same fit but 
      where the crust has been replaced with the one of the original barotropic RG(SLy4)~\citep{gulminelli_unified_2015, chabanat_skyrme_1998, danielewicz_symmetry_2009} table.}
    \label{fig:mr_crustinfluence}
  \end{figure}

\section{Representation of two-argument equations of state}\label{sec:twoparameos}

For the purposes of cold EoSs that describe matter out of the weak $\beta$-equilibrium, we shall
consider General Purpose tables from \textsc{CompOSE} that are
described with the three arguments $(n_B,\,Y_e,\,T)$ and, as an
approximation, use the first temperature entry of the table, which in
general corresponds to a temperature $T\lesssim100$ keV. The thermal effects at
this temperature are relevant only in the outer part of the outer
crust. This part of the EoS is not captured by the fits, therefore we
consider that the first temperature entry of the table is an
  excellent approximation to the zero-temperature case.

{We will adapt the procedure described in the previous section for the two-argument form of the tables. In addition to imposing pressure and sound speed continuity, we also ensure the existence of $\beta$-equilibrium for all densities.} The neutrinoless $\beta$-equilibrium condition is:
\begin{equation}\label{eq:mulequalszero}
	\mu_{l_e} = 0,
\end{equation}
where $\mu_{l_e}$ is the chemical potential of leptons. Using the definition $\mu_{l_e}=\left(\drond[\epsilon]{n_e}\right)_{n_B}$ (see Sec.~3.7 of~\citep{servignat_new_2023}), it can be rewritten as:
\begin{equation}
    \left(\drond[\epsilon]{Y_e}\right)_{n_B} = 0,
\end{equation}
{implying} that $\forall n_B$, $\epsilon$ {must} have a minimum in the $Y_e$ direction. Because the original EoS should also possess a $\beta$-equilibrium, this condition should automatically be fulfilled in the high-density part, provided that the fit is accurate enough. However, as the low-density part {consists of} a generic polytropic crust, special care should be brought when constructing it for EoSs with two arguments.

The procedure is the following: 
according to the frozen composition principle, the two-argument EoS is considered as a collection of $N_Y$ one-argument EoSs, $N_Y$ being the number of tabulated values of $Y_e$, each of which is fitted according to the procedure described in Sec.~\ref{sec:oneparameos}, meaning that we once again have three parts: a high-density part where the fit is performed, a low-density part where a simplified polytropic model is applied, and an intermediate part described with a single-piece GPP to connect the two other parts. This time, for the pseudo-polytrope
we choose to {perform the} fit on $X=\{\epsilon/n_B\}$ only, to help reduce parameter dispersion from one slice to the next.
{For the Potekhin-Pearson scheme, we perform a fit of the pressure only.}
Each new fit is initialized with the optimal coefficients of the previous one.

For each fit, the outermost polytrope has $\gamma$ fixed and $\kappa$ taken to be a polynomial in $Y_e$:
\begin{align}
	\kappa(Y_e) & = \sum\limits_{i=0}^{n_\kappa}\kappa_iY_e^i.
\end{align}
We determine the coefficients $\kappa_i$ as a polynomial fit to $\epsilon(Y_e, n_{\mathrm{lim},2})$, i.e. we write $\forall i\in\llbracket 0,\,n_\kappa\rrbracket,\, \kappa_i=A\kappa'_i$ and determine the $\{\kappa'_i\}_{i\in\llbracket 0,\,n_\kappa\rrbracket}$ with the fit. Then the constant $A$ is chosen freely. This procedure guarantees that there is always a $\beta$-equilibrium solution in the low-density part. Only the intermediate part is left, where GPP expressions are used to connect the crust to the fit. They are built to provide pressure and sound speed continuity at the junctions, but there is no simple theoretical argument that ensures the existence of a $\beta$-equilibrium in the intermediate part. However, we find that at the cost of fine-tuning the parameter $A$ by hand, this is always the case. The result of the fitting procedure is a collection of $N_Y$ coefficient lists.
For example with the pseudo-polytrope, using 7 coefficients in Eq.~\eqref{eq:coefsg} gives 16 fitting coefficients per slice: 8 + 2 for the high-density region and 6 that are deduced in the mid-density region to ensure that the matching conditions are fulfilled. In the end, the formalism is semi-analytic because to compute the thermodynamics at a non-tabulated value of $Y_e$, one must interpolate between two neighbor fit expressions. Moreover, producing initial data requires the computation of the $\beta$-equilibrated EoS, which is entirely numerical. 
For example, the $\beta$-equilibrated EoS yielded by a two-argument pseudo-polytrope is not a one-argument pseudo-polytrope, but is rather a curve in the $(n_B, Y_e)$ plane. 
{It is important to note that the procedure makes no assumption regarding the fitting scheme}, which means that in principle it can be applied to any fit of the high density part of a given EoS.

\begin{figure*}
  \begin{minipage}[t]{0.95\textwidth}
  \begin{subfigure}[b]{0.45\textwidth}
    \centering
    \includegraphics[width=\textwidth]{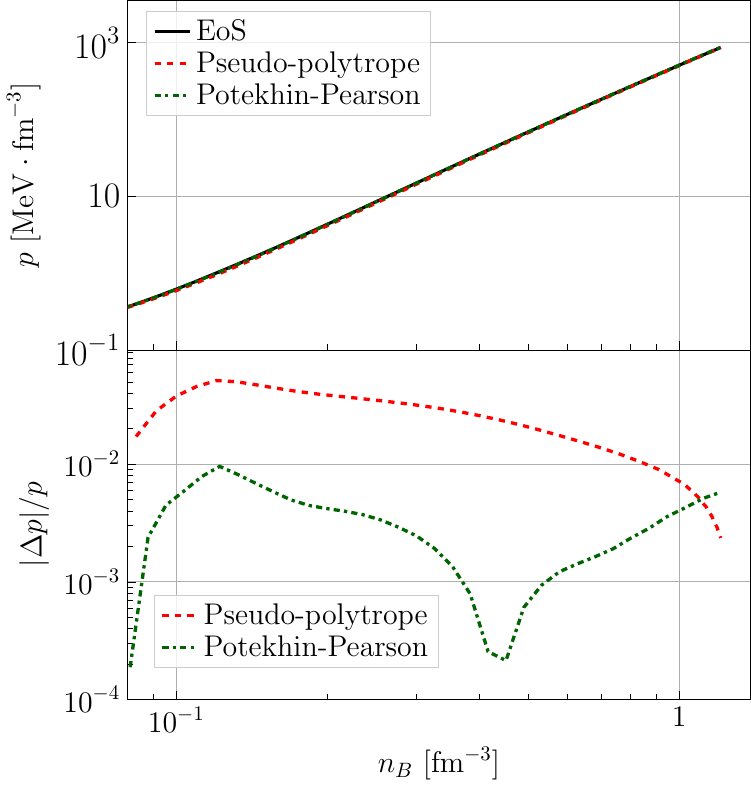}
  \end{subfigure}
  \hfill
  \begin{subfigure}[b]{0.45\textwidth}
    \centering
    \includegraphics[width=\textwidth]{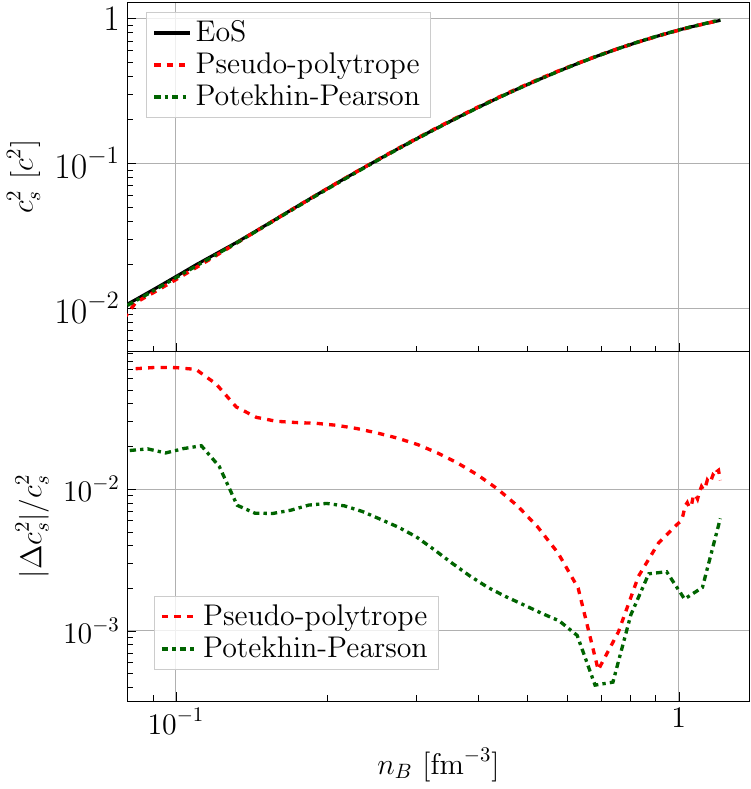}
  \end{subfigure}
  \end{minipage}
  \begin{minipage}[b]{0.95\textwidth}
  \begin{subfigure}[b]{0.45\textwidth}
    \centering
    \includegraphics[width=\textwidth]{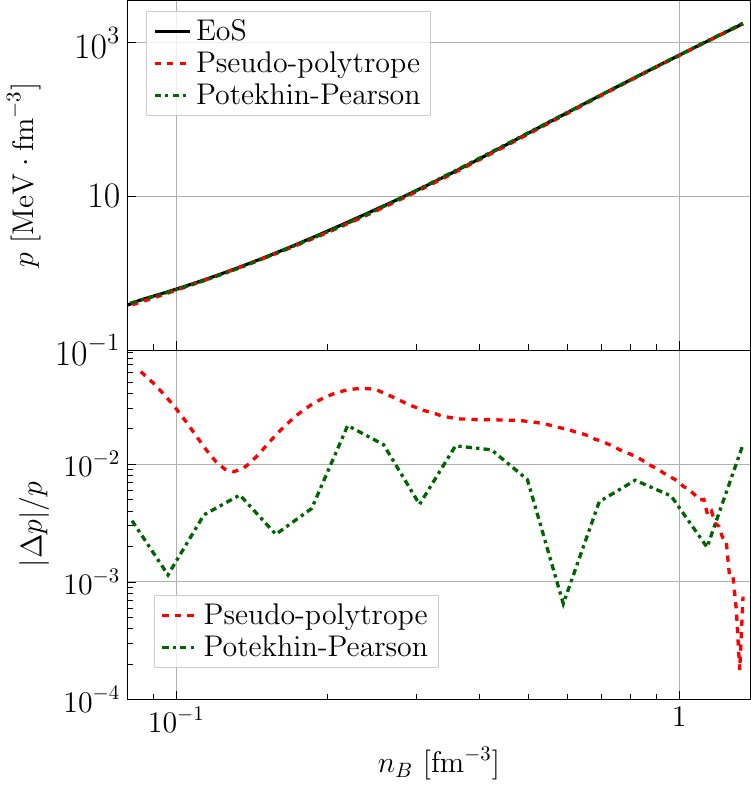}
    \caption{Pressure vs baryon number density.}
  \end{subfigure}
  \hfill
  \begin{subfigure}[b]{0.45\textwidth}
    \centering
    \includegraphics[width=\textwidth]{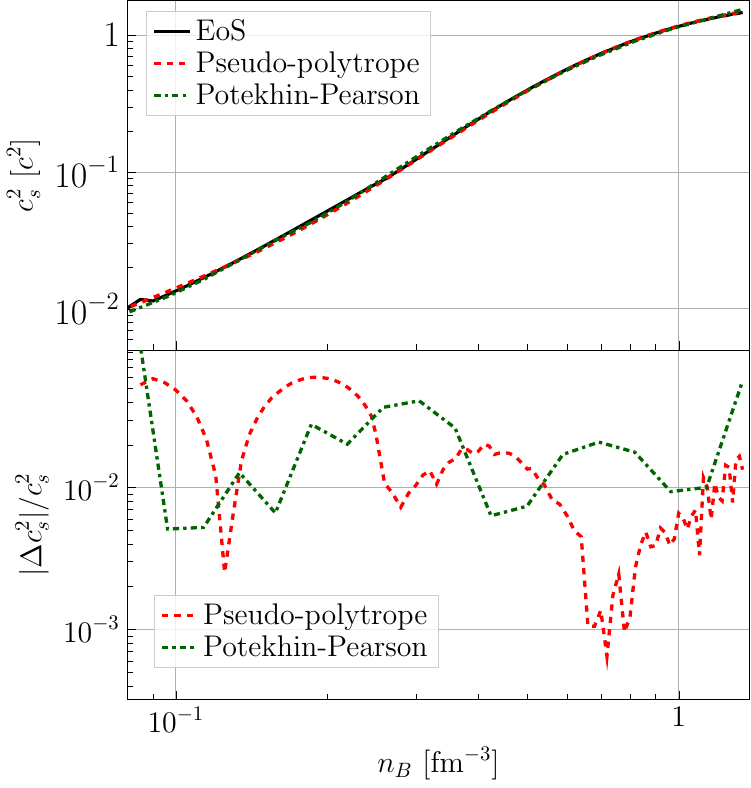}
    \caption{Sound speed vs baryon number density.}
  \end{subfigure}
  \caption{Thermodynamical profiles of two barotropic EoSs: RG(SLy4)~\citep{gulminelli_unified_2015, chabanat_skyrme_1998, danielewicz_symmetry_2009} (top) - APR(APR)~\citep{haensel_neutron_2007, akmal_equation_1998} (bottom).}
  \label{fig:thermosly4apr-1parameter}
  \end{minipage}
  \end{figure*}
  
\begin{figure*}
\begin{minipage}[t]{0.95\textwidth}
\begin{subfigure}[b]{0.45\textwidth}
	\centering
	\includegraphics[width=\textwidth]{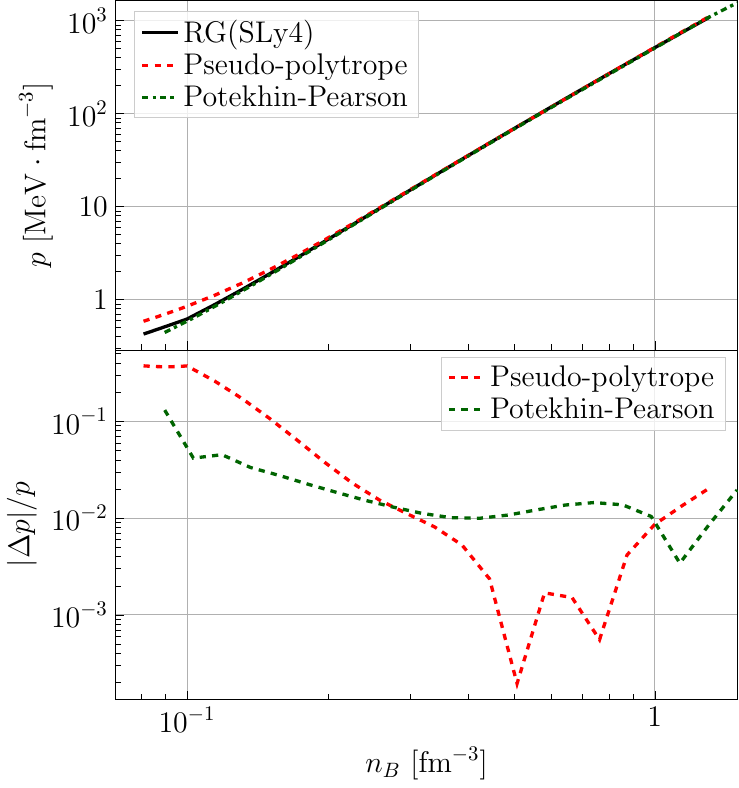}
\end{subfigure}
\hfill
\begin{subfigure}[b]{0.45\textwidth}
	\centering
	\includegraphics[width=\textwidth]{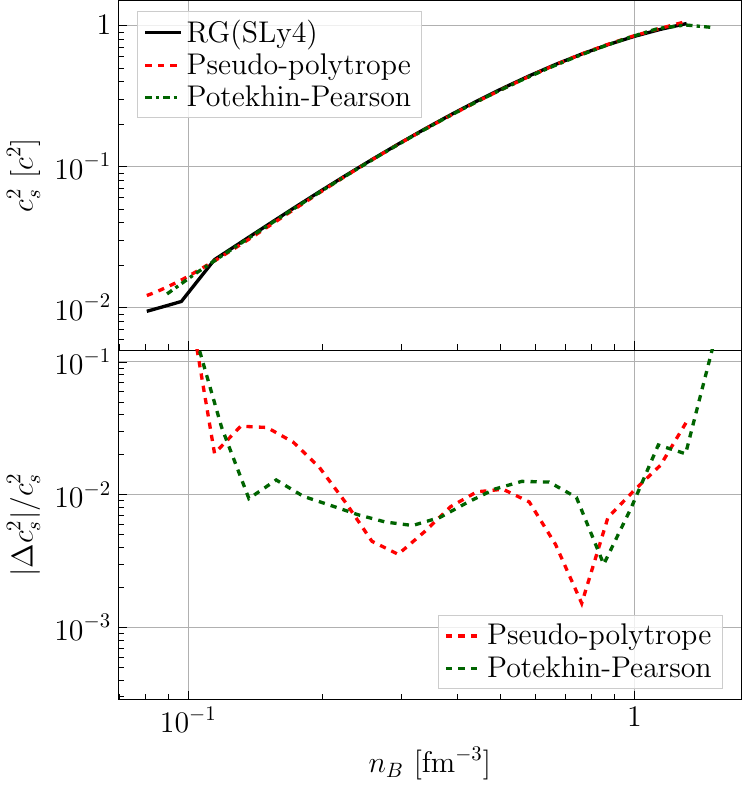}
\end{subfigure}
\end{minipage}
\begin{minipage}{0.95\textwidth}
\begin{subfigure}[b]{0.45\textwidth}
	\centering
	\includegraphics[width=\textwidth]{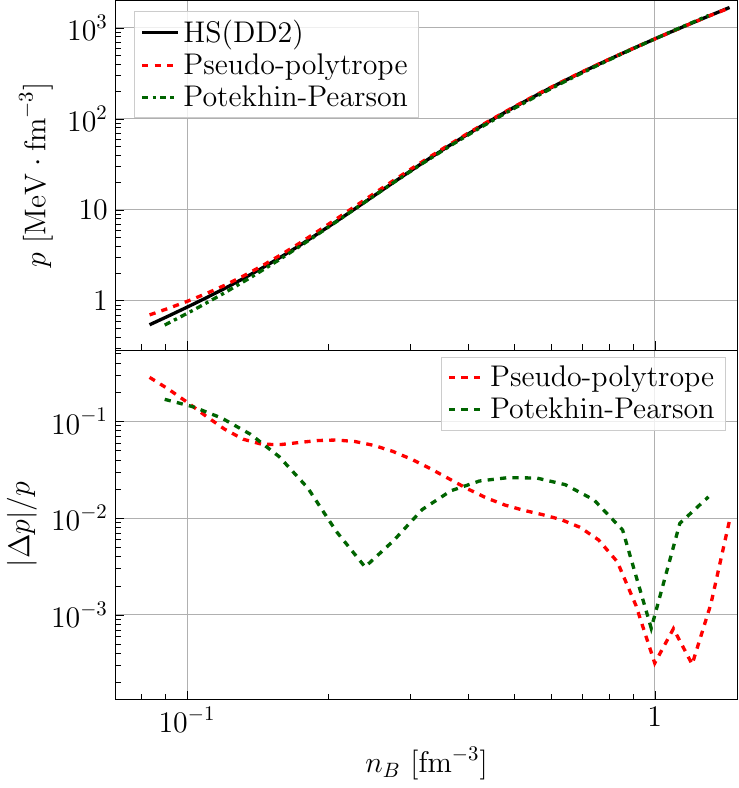}
\end{subfigure}
\hfill
\begin{subfigure}[b]{0.45\textwidth}
	\centering
	\includegraphics[width=\textwidth]{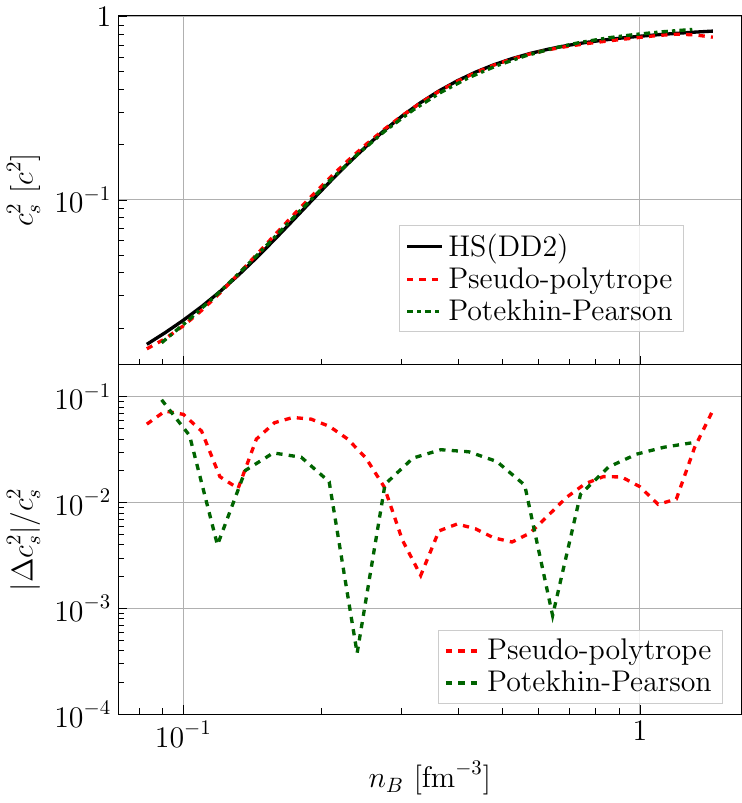}
\end{subfigure}
\end{minipage}
\begin{minipage}[b]{0.95\textwidth}
\begin{subfigure}[b]{0.45\textwidth}
	\centering
	\includegraphics[width=\textwidth]{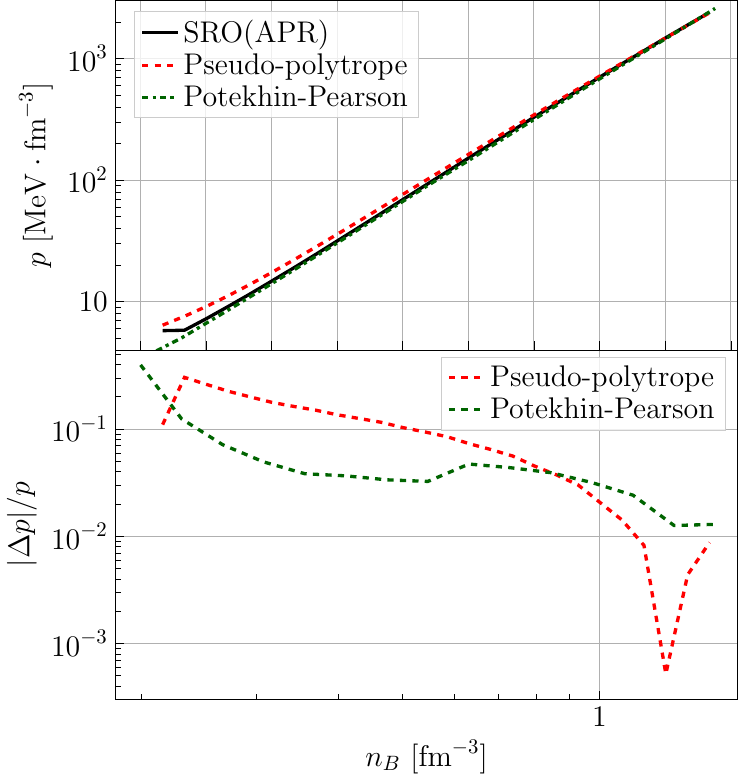}
	\caption{Pressure vs baryon number density.}
\end{subfigure}
\hfill
\begin{subfigure}[b]{0.45\textwidth}
	\centering
	\includegraphics[width=\textwidth]{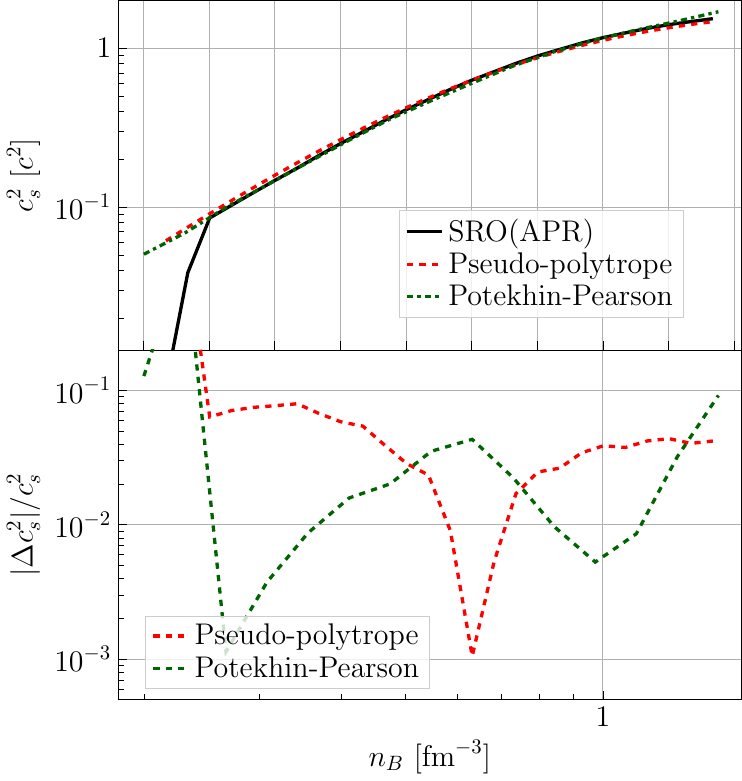}
	\caption{Sound speed vs baryon number density.}
\end{subfigure}
\caption{Two-argument fits: comparison between the $\beta$-equilibrated versions of the original tables and the $\beta$-equilibrated versions of the fits.}
\label{fig:thermosly4apr-2parameter}
\end{minipage}
\end{figure*}

\section{Results}\label{sec:results}

In this section all tests are run with isolated, spherically symmetric, non-rotating NSs using the code described in~\citep{servignat_new_2023}. The fits are tested on the EoSs described in the introduction.

\subsection{Static quantities}
In order to check our fits' performances we first check the static quantities: mass $M$, radius $R$, tidal deformability $\Lambda$. The definition of the tidal deformability is the following~\citep{hinderer_tidal_2008, malik_constraining_2018}:
\begin{equation}
	\Lambda = \frac{2}{3}\Xi^{-5}k_2,
\end{equation}
where $\Xi=M/R$ is the compactness parameter and $k_2$ the so-called
$\ell=2$ tidal Love number~\citep{hinderer_tidal_2008}. These
quantities can be obtained from an $\ell=2$ linear perturbation of a
spherically symmetric star equilibrium obtained from the
Tolmann-Oppenheimer-Volkoff equations~\citep{hinderer_tidal_2008,
  hinderer_tidal_2010}. To compare the static quantities, we compute
the mass-deformability diagrams of the fitted EoS and we compare them
with the curves obtained with the original table. The results are
shown on Figs.~\ref{fig:lmdiagSLy4} and~\ref{fig:lmdiagAPR} for the
one-argument fits, and Figs.~\ref{fig:lmdiagDD2},
~\ref{fig:lmdiagSROAPR} and~\ref{fig:lmdiagRGSLy4} for the
two-argument fits. The fits capture well the deformability: regarding
the one-argument EoSs the error on the computation of
$\Lambda$ {is} 3\% for RG(SLy4)~\citep{gulminelli_unified_2015,
  chabanat_skyrme_1998, danielewicz_symmetry_2009} and 4\% for
APR(APR)~\citep{haensel_neutron_2007, akmal_equation_1998} with the
pseudo-polytrope, and the Potekhin-Pearson fit reproduces the values of
$\Lambda$ with an error that never exceeds 1\%. The fits perform
well for two-argument EoSs as the error is also of the order of a few
percent, except for the
SRO(APR) EoS~\citep{schneider_open-source_2017,schneider_akmal_2019}
{for which the phase transition to a pion condensate around
  $n_B=0.2\,\mathrm{fm}^{-3}$ is treated with a Maxwell type
  construction}, which is not well-reproduced by our fits, and where
the error of the pseudo-polytrope exceeds 10\%. 
The
Potekhin-Pearson approach does slightly better 
but the error remains close to
10\% {as again the phase transition region is not well reproduced}. For SRO(APR) we point out that the minimization procedure to add the crust,
as described in Section~\ref{subsec:pseudopoly}, fails to converge for the Potekhin-Pearson scheme if we use $n_{\mathrm{lim,2}}=0.2$ fm$^{-3}$ as applied by the
pseudo-polytrope formalism. To remove the phase transition associated with the pions, we instead replace the tabulated {values of the adiabatic index, $\Gamma=\mathrm{d}\mathrm{log}p/\mathrm{d}\mathrm{log}e$ in the vicinity of the phase transition with smoothed values using linear interpolation. The pressure is then recomputed from these updated values of the adiabtic index.} Note that
the one-argument APR(APR) EoS considers a mixed phase at the
transition to the pion-condensed phase which smoothens the EoS and
that it can thus be much better reproduced by our fits.

These figures also show that the maximum masses for each of the EoSs considered are
well-reproduced (see Table~\ref{tab:comparisonparametrizations} of App.~\ref{s:fit_coefs}). On
the other hand, we know that the predicted radius strongly depends on
how the crust-core transition is made, as well as the actual crust
model~\citep{fortin_neutron_2016,suleiman_influence_2021}; the
error can reach one kilometer { for a 1$\Msol$ neutron star. We emphasize that changing the matching between the crust and the core may induce this error even without changing the physics.} The drastic effect of the
crust on the radius is shown in Fig.~\ref{fig:mr_crustinfluence},
where the mass radius diagrams of the barotropic RG(SLy4)
EoS~\citep{gulminelli_unified_2015, chabanat_skyrme_1998,
  danielewicz_symmetry_2009} is compared with the pseudo-polytropic
fit, as well as the fit where the EoS crust was replaced with the one
of the original EoS.
{For a canonical $1.40\Msol$ neutron star, we compare the radii and deformabilities with stars constructed with the RG(SLy4) EoS and its fits: (i) the original table, (ii) the pseudo-polytropic fit and (iii) the pseudo-polytropic fit where the crust has been replaced with the one of the original table. We give the values for RG(SLy4):
\begin{enumerate}[(i)]
	\item $R=11.7$ km and $\Lambda=297$
	\item $R=11.5$ km and $\Lambda=290$
	\item $R=11.9$ km and $\Lambda=290$
\end{enumerate}
and APR(APR):
\begin{enumerate}[(i)]
	\item $R=11.3$ km and $\Lambda=249$
	\item $R=11.2$ km and $\Lambda=240$
	\item $R=11.2$ km and $\Lambda=231$
\end{enumerate}
Replacing the crust has changed the value of the radius and the deformability by a few percents while Fig.~\ref{fig:crustinfluencefreqs} shows that the frequencies remain unchanged when varying the crust model.}
The pressure and sound speed profiles are
represented in Fig.~\ref{fig:thermosly4apr-1parameter} for the
one-argument EoSs and Fig.~\ref{fig:thermosly4apr-2parameter} for the
$\beta$-equilibrated version of the two-argument EoSs. All in all
the fits show a good agreement compared to the original EoS,
especially when no phase transition is present, as the relative difference in the thermodynamic quantities
between the fit and the original EoS remains below 10\%, except close to $n_{\mathrm{lim},2}$ where the low-density matching
procedure tends to mildly degrade the quality of the fit.
The phase transition of
SRO(APR)~\citep{schneider_open-source_2017,schneider_akmal_2019} can
be seen in Fig.~\ref{fig:thermosly4apr-2parameter}: the sudden drop of
the sound speed is due to a discontinuity in the pressure derivative.
 
\subsection{Dynamic quantities}
\begin{figure}
\centering
\includegraphics[width=0.4\textwidth]{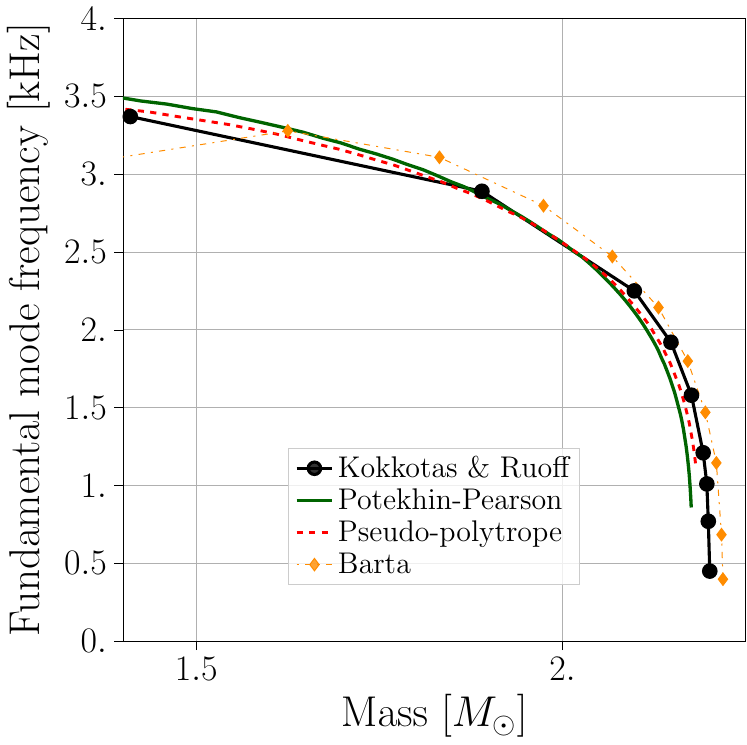}
\caption{$f(M)$ diagram for barotropic APR(APR)~\citep{haensel_neutron_2007, akmal_equation_1998} EoS compared with values taken from Table A.17 of~\citep{kokkotas_radial_2001} and Table 2 of~\citep{barta_fundamental_2021}.}
\label{fig:massfreqAPRkokkotasruoff}
\end{figure}

\begin{figure}
\centering
\includegraphics[width=0.4\textwidth]{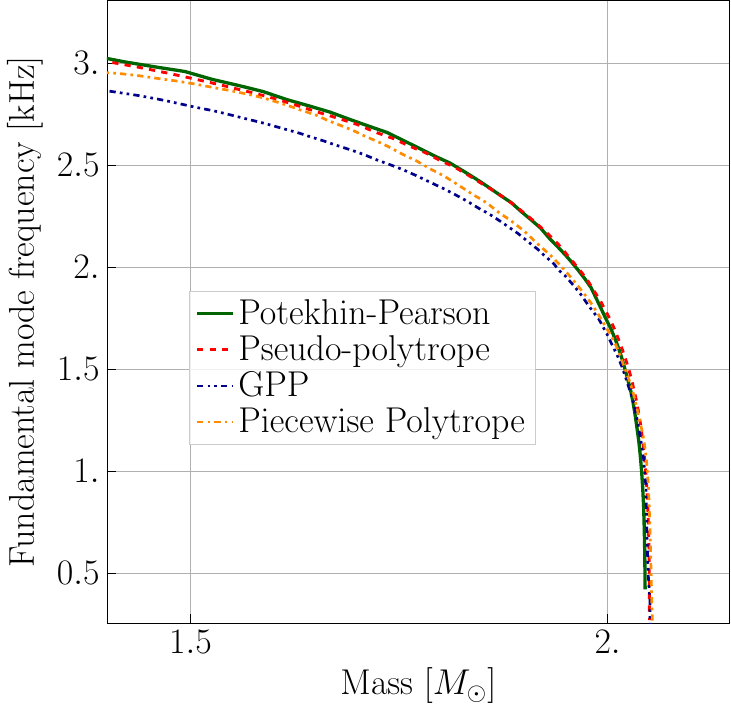}
\caption{$f(M)$ diagram for barotropic RG(SLy4)~\citep{gulminelli_unified_2015, chabanat_skyrme_1998, danielewicz_symmetry_2009} EoS. The data for the GPPs and Piecewise Polytropes have been digitized from~\citep{oboyle_parametrized_2020} with \texttt{engauge-digitizer}~\citep{mitchell_engauge_2023}.}
\label{fig:massfreqsly4}
\end{figure}

We also compare
the frequencies yielded by using the fits in our code that was presented and benchmarked in~\citep{servignat_new_2023} with
those tabulated in the literature. The frequency extraction
procedure is a maximum search in the spectrum coupled with a quadratic
interpolation of the closest neighbors. As shown on
Fig.~\ref{fig:massfreqAPRkokkotasruoff}, our frequencies are in good
agreement with the ones tabulated in~\citep{kokkotas_radial_2001,
  barta_fundamental_2021}, where a perturbative approach is used.
  {Although Barta~\cite{barta_fundamental_2021} has developed a dissipative formalism
  in the perturbative framework, his approach was benchmarked on the EoSs of~\citep{kokkotas_radial_2001}
  and the EoSs that are used in those two papers are the same versions of APR and SLy4 chosen here.}
  The error of the fits on the sound speed profile is comparable with the error between the approaches of~\citep{kokkotas_radial_2001}
  and~\citep{barta_fundamental_2021}, we therefore consider the values of frequencies to be compatible with one another.
Fig.~\ref{fig:massfreqsly4} shows the mass-frequency diagram for
the RG(SLy4) EoS~\citep{gulminelli_unified_2015, chabanat_skyrme_1998,
  danielewicz_symmetry_2009}. Even though in
Ref.~\citep{barta_fundamental_2021} frequency values for the
fundamental mode of oscillating spherically symmetric NS are presented
for this particular EoS, those frequencies do not vanish as the
mass approaches the maximum mass of the EoS. This suggests that there was
an issue in the computation of the frequencies and we therefore do not
include them in the figure. For both $f(M)$ diagrams the
frequency of the fundamental mode approaches zero {towards} the
maximum mass as expected. It is also notable that the Potekhin-Pearson approach and the
pseudo-polytrope give very close frequencies one to
another. Considering that the Potekhin-Pearson approach uses 15 fit coefficients
per slice while the pseudo-polytropic approach uses only 3 to 7 fit
coefficients per slice, the latter shows its efficiency in reproducing
dynamical NS oscillation modes compared to the former.

\begin{figure}
	\centering
	\includegraphics[width=0.4\textwidth]{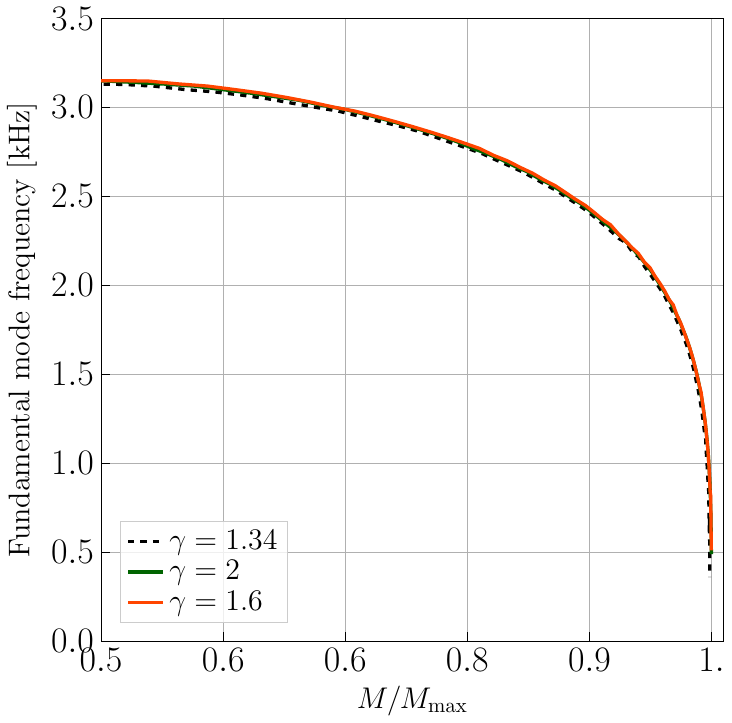}
	\caption{$f(M)$ diagram for three pseudo-polytropic fits with three different choices of polytropic index in the crust.}
	\label{fig:crustinfluencefreqs}
\end{figure}

The influence of the crust on the frequencies is shown in Fig.~\ref{fig:crustinfluencefreqs}: a pseudo-polytropic fit is performed for three different values of the polytropic index in the crust and the $f(M)$ diagram computed for the three fits. The $x$-axis of the plot is the mass rescaled to the maximum mass yielded by the fit to aid comparison. The frequencies are almost unchanged between the three fits, which is in contrast with the influence the choice of the crust has on the radius. It shows the limited influence of the crust dynamics on the oscillation modes of a NS. A possible explanation for this is that we do not actually take the detailed physics of the crust into account in the hydrodynamic simulations as the whole star is modeled by a perfect fluid,
and an accurate description of the crust would correspond to use models of the cristalline structure. Also at this stage the hydrodynamical code does not support the non-monotonic behavior of the sound speed in the crust of~\citep{douchin_unified_2001}.

\begin{figure*}
  \begin{minipage}[t]{.95\textwidth}
    \begin{subfigure}[b]{.327\textwidth}
    \centering
    \includegraphics[width=\columnwidth]{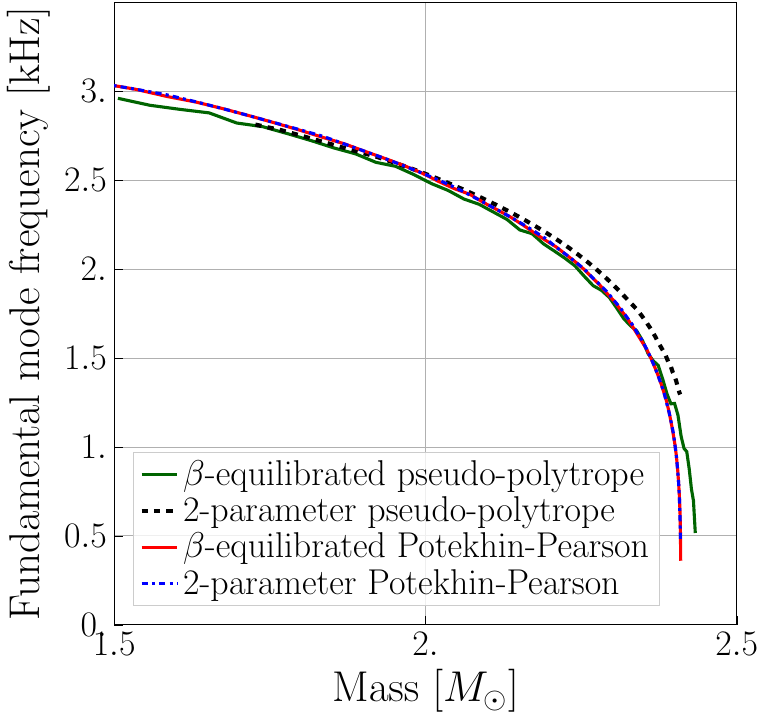}
    \caption{HS(DD2)~\citep{hempel_statistical_2010,typel_composition_2010}.}
    \label{fig:mfdiagdd2}
    \end{subfigure}
    \hfill
    \begin{subfigure}[b]{.29\textwidth}
    \centering
    \includegraphics[width=\columnwidth]{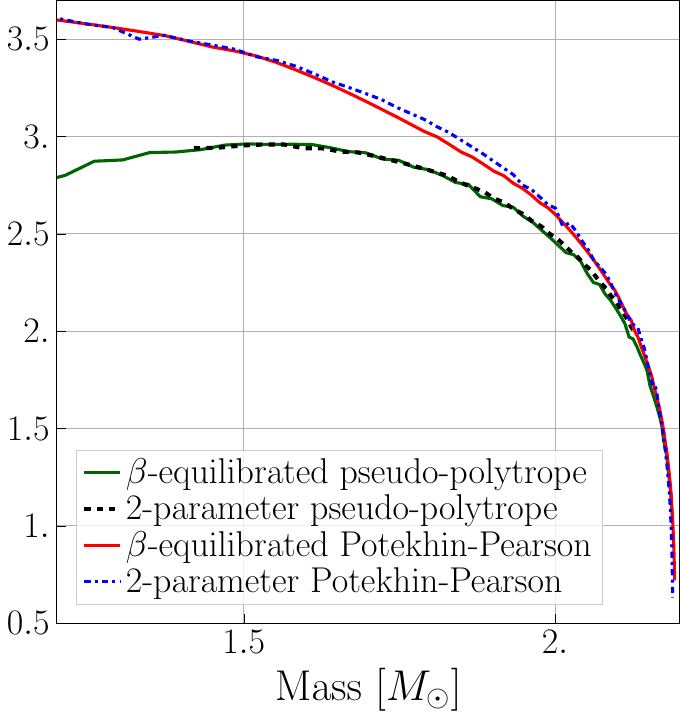}
    \caption{SRO(APR)~\citep{schneider_open-source_2017,schneider_akmal_2019}.}
    \label{fig:mfdiagapr}
    \end{subfigure}
    \hfill
    \begin{subfigure}[b]{.29\textwidth}
    \centering
    \includegraphics[width=\columnwidth]{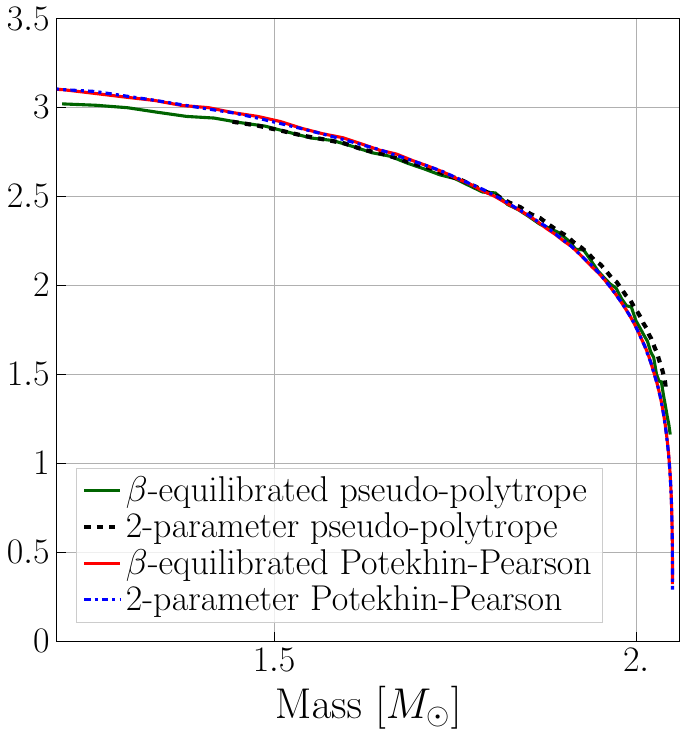}
    \caption{RG(SLy4)~\citep{gulminelli_unified_2015, raduta_nuclear_2019}.}
    \label{fig:mfdiagsly4}
    \end{subfigure}
  \end{minipage}
  \caption{$f(M)$ diagram for the three General Purpose EoSs.}
  \label{fig:mfdiagrealistic}
  \end{figure*}

\begin{figure}
	\centering
	\includegraphics[width=0.4\textwidth]{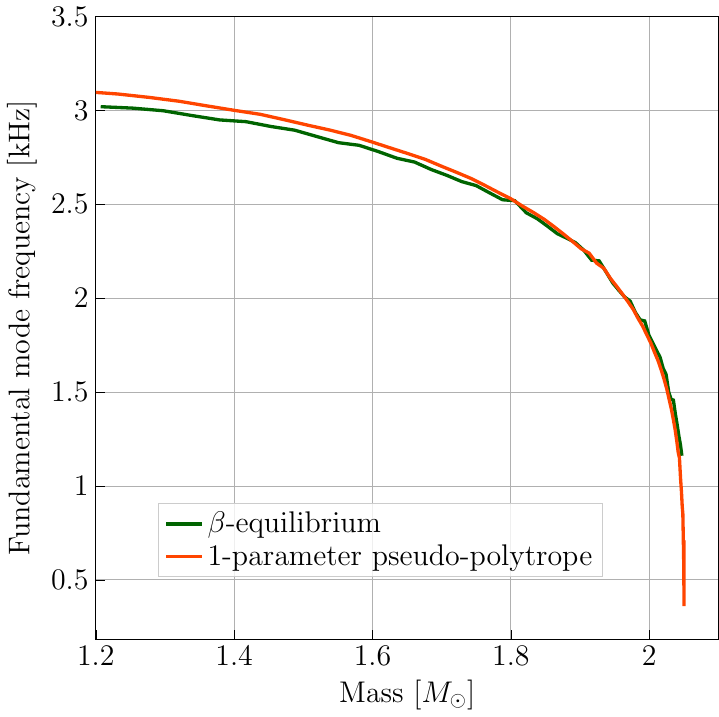}
	\caption{Comparison of fundamental mode frequency between the one-argument pseudo-polytropic fit of barotropic RG(SLy4)~\citep{gulminelli_unified_2015, chabanat_skyrme_1998, danielewicz_symmetry_2009} and the $\beta$-equilibrated version of the two-argument pseudo-polytropic fit of RG(SLy4)~\citep{gulminelli_unified_2015, raduta_nuclear_2019}.}
	\label{fig:comparaison1Dfitbetaeq}
\end{figure}

The $f(M)$ curves for the two-argument fits of the three chosen
General Purpose EoSs are plotted on
Fig.~\ref{fig:mfdiagrealistic}. The frequencies agree quite well
between the two fitting procedures, and it is also notable that going
from the $\beta$-equilibrated version of the EoS to the two-argument
version gives very similar frequencies.
We note that, at low neutron star masses, the Potekhin-Pearson scheme predicts higher fundamental frequencies than the pseudo-polytrope formalism.
This could be a consequence of the impact that the different values of $n_{\mathrm{lim,2}}$ used by the two fitting schemes has on the properties of
the intermediate region that connects the polytropic crust with the fitted core. Therefore, for SRO(APR), the intermediate region may have a non-neglible
impact on the fundamental frequencies.
Unlike the EoSs considered in Fig. \ref{fig:comparaison1Dfitbetaeq}, the $f(M)$ curves for the $\beta$-equilibrated and two-argument versions of SRO(APR) do not match.
This could be a consequence of the crude smoothing used to remove the pion condensate.
The fundamental difference
between the two simulations is the production of neutrinos, which is
here taken into account through the evolution equation for $Y_e$.
This is a simple advection equation with a source term, see App.~\ref{app:evolutionelectronfraction}. 
The source terms
$\sigma$ are computed from the expressions
in~\citep{haensel_non-equilibrium_1992}, and depend strongly on the
temperature, namely being proportional to $(T/10^9\mathrm{K})^5$ for
direct Urca processes and $(T/10^9\mathrm{K})^7$ for modified Urca
processes. Because we choose the first entries of the tables that
typically correspond to a temperature of 100 keV, the actual
computation of source terms gives values so small that they are below
the machine accuracy and therefore are virtually zero. Therefore the simulations
happen exactly as if no neutrinos were emitted. The initial data used to perform the evolution are computed
with the $\beta$-equilibrated version of the fits and then the star numerically exits
the $\beta$-equilibrium state thanks to the fact that $\mu_{l_e}\neq0$ numerically.

\begin{figure}
	\centering
	\includegraphics[width=0.4\textwidth]{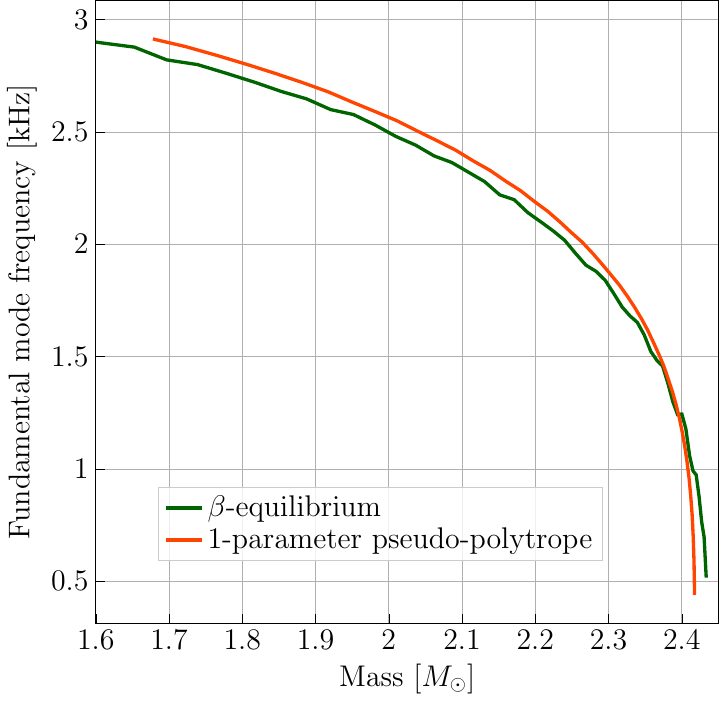}
	\caption{Comparison of fundamental mode frequency between the one-argument pseudo-polytropic fit of barotropic GPPVA(DD2)~\citep{typel_composition_2010,grill_equation_2014,pearson_unified_2018} and the $\beta$-equilibrated version of the two-argument pseudo-polytropic fit of HS(DD2)~\citep{hempel_statistical_2010,typel_composition_2010}.}
	\label{fig:comparaison1Dfitbetaeqbis}
\end{figure}

Comparing the results of the one- and two-argument fits
for RG(SLy4)~\citep{gulminelli_unified_2015, chabanat_skyrme_1998,
  danielewicz_symmetry_2009} shows that the frequencies 
converge for higher masses,
cf. Figs.~\ref{fig:comparaison1Dfitbetaeq} and~\ref{fig:comparaison1Dfitbetaeqbis}. The small discrepancy for higher mass is due to the slightly different value of maximum mass
between the $\beta$-equilibrated version of the two-argument fit of HS(DD2)~\citep{hempel_statistical_2010,typel_composition_2010} and the one-argument fit of GPPVA(DD2)~\citep{typel_composition_2010,grill_equation_2014,pearson_unified_2018}.
This is consistent with the previous paragraph.
We recall that the tables
are slightly different: the one-argument versions on \textsc{CompOSE}
exactly correspond to zero-temperature whereas the General Purpose
tables we used for two-argument fits start at a low but non-zero
temperature. The latter also rely on more general calculations that do
not consider a cristalline structure in the low-density inhomogeneous
phase. However, as the study of the influence of the crust on the
frequencies of the fundamental mode suggests, these differences that
mainly concern the crust should have very little effect on the values
of the frequencies.
Also, the direct comparison cannot be made for APR, because even
though~\citep{schneider_open-source_2017} uses the same nuclear
interaction as~\citep{akmal_equation_1998}, the final EoS is
different: the one-argument version was computed with a mixed phase
whereas the general purpose one was computed with two distinct phases,
yielding a first order transition. {Overall}, the differences are most certainly due to the
increased number of fit coefficients for the two-argument functional.

\section{Conclusions}\label{sec:conclusion}
We presented two systematic ways to represent any given EoS by an
analytic fit function. The low-density (crust) part of the EoSs is not
represented by the fits and we consider simplified expressions for
this part. In principle, we could match a widely used crust, for
instance the one from \citep{douchin_unified_2001}, or attach a
consistent crust obtained e.g. from the \textsc{Cuter}
tool~\citep{davis_cuter_2023}. This tool, based on the work of
\cite{Carreau:2019zdy}, allows to attach a physically accurate crust to
any EoS. The hydrodynamical code that we use, however, does not
support the non-monotonic behavior of the sound speed that happens
around the transition from the inner to the outer crust at $n_B\approx
2\times10^{-4}\,\mathrm{fm}^{-3}$. This is also the reason why we drop
the crust terms in the Potekhin-Pearson model. These issues might
be related to the non-convex hydrodynamics triggered by such
behaviors, in connection with the so-called fundamental
derivative~\citep{ibanez_convexity_2013, ibanez_anomalous_2018}. Our simplfied
approach can be justified by the fact that the detailed crust physics
is not the most important when looking at NS oscillations.

A first asset of the approach lies in its economical
nature: instead of storing a large table, any EoS can be represented in
the form of a few coefficients and a formula, as an be expected from a
parametrization. For example, the General Purpose tables used in the
paper have a total of 826 708
(RG(SLy4)~\citep{gulminelli_unified_2015, raduta_nuclear_2019}), 1 239
300 (HS(DD2)~\citep{hempel_statistical_2010,typel_composition_2010})
and 3 151 302
(SRO(APR)~\citep{schneider_open-source_2017,schneider_akmal_2019})
thermodynamic entries (the files’ sizes are respectively 70, 167 and 496 MB.
The number of data entries reduce to 6 667, 15 300, and 23 694, respectively, when considering only the lowest temperature entry). On
the other hand fitting with the Potekhin-Pearson model gives 15 coefficients
per slice which corresponds to around 900 coefficients, and the pseudo
polytrope gives 3 to 7 coefficients per slice, depending on the choice
of the degree of the polynomial $g$, which amounts to 200 to 500
coefficients. A second asset is the low-density stitching procedure:
considering an analytic low-density model, it gives continuous sound speed profiles {even down to zero densities}, which is
important for dynamical simulations of neutron stars, and the procedure is in
principle applicable to any EoS parametrization. {Finally, we have demonstrated that the approach made the fitting of two-argument EoSs possible thanks to the simplified crust model, from which we were then able to compute macroscopic static quantities as well as to perform dynamical simulations from which frequencies could be extracted.} The Potekhin-Pearson
model allowed to compare our novel approach with an already
existing fitting scheme that is known for its precision in the
thermodynamical profiles. The pseudo-polytrope does almost as well as
the Potekhin-Pearson approach in reproducing the macroscopic static quantities
of the original EoS, except in the presence of a phase transition
where the thermodynamics is harder to capture, but where both
approaches fail to reproduce it faithfully. However the fundamental
radial frequencies of one-argument EoSs are well reproduced by both approaches, and compatible when independently applied to
two-argument EoSs. In that regard, the pseudo-polytrope might be
preferable as the number of fitting coefficients is reduced compared
to the Potekhin-Pearson approach. We provided, whenever possible, a comparison between 
the published values of macroscopic static quantities and radial oscillation frequencies obtained with previous EoS parametrizations.
We also have shown
that changing the crust has a very limited effect on the values of the oscillation frequencies of neutron stars,
especially for stars with a high mass where the crust is thinner.
We recall here that we do not use a detailed description of the crust in the hydrodynamical simulations and further
work in that direction would be needed to assess the extent of this result.
Extensions of
the work would be improving the two-argument fitting scheme by making
it fully analytical, and extending it to three-argument EoS tables
with temperature as an additional argument. 
We give a few words on EoSs that include hyperons: we tried to apply the fits on one of those EoSs. Both fitting procedures were unable to successfully capture the phase transition at high densities which induces a discontinuity in the sound speed. Since the discontinuity is physical, one way to address this would be to consider several zones with different fitting functionals. This would be worth exploring for future versions of the fits.
We plan to make the
pseudo-polytrope fitting code
publicly available as a tool of
\textsc{CompOSE} in the future.

\section*{Data availability statement}
The codes used in this study are still in developement and have not been published yet. The data that support the findings of this study are available upon reasonable request from the authors.

\section*{Acknowledgements}
GS would like to thank Michael O'Boyle for his useful insight on the
Generalized Piecewise Polytropes and Dániel Barta for details on his
paper on radial oscillations of neutron stars. GS, PJD, JN and MO
acknowledge financial support from the Agence Nationale de la
Recherche (ANR) under contract ANR-22-CE31-0001-01 and from the CNRS International Research Project (IRP)
“Origine des éléments lourds dans l’univers: Astres Compacts et Nucléosynthèse (ACNu)”. JAP acknowledges
the support through the grant PID2021-127495NB-I00 funded by
MCIN/AEI/10.13039/501100011033 and by the European Union, the
Astrophysics and High Energy Physics programme of the Generalitat
Valenciana ASFAE/2022/026 funded by MCIN and the European Union
NextGenerationEU (PRTR-C17.I1) and the Prometeo excellence programme
grant CIPROM/2022/13. The authors gratefully acknowledge the Italian
Instituto Nazionale de Fisica Nucleare (INFN), the French Centre
National de la Recherche Scientifique (CNRS) and the Netherlands
Organization for Scientific Research for the construction and
operation of the Virgo detector and the creation and support of the
EGO consortium.

\bibliographystyle{apsrev-nourl}
\bibliography{biblio}

\appendix
\section{Analytical formul{\ae} of pseudo-polytropic thermodynamics}
The internal energy $\epsilon$
serves as a potential
from which all other thermodynamic quantities can be derived, and we
here compile the expressions of the most common ones, expressed as a
function of $\epsilon$ and its derivatives or as the fitting
coefficients of the expression~\eqref{eq:pseudopolytrope} and
$x=\ln(n_B [\mathrm{fm}^{-3}])$:
\begin{align}
	\epsilon(x) & = e^{\alpha x}\sum\limits_{k=0}^n\bar{a}_kx^k + d - e^{-x}L \\
	\deriv[\epsilon]{x}(x) & = e^{\alpha x}\left[\alpha\sum\limits_{k=0}^n\bar{a}_kx^k + \sum\limits_{k=1}^nk\bar{a}_kx^{k-1}\right] + e^{-x}L \\
	\derivdeux[\epsilon]{x}(x) & = e^{\alpha x}\left[\alpha^2\sum\limits_{k=0}^n\bar{a}_kx^k + 2\alpha\sum\limits_{k=1}^nk\bar{a}_kx^{k-1}\right. \nonumber \\
							   & \left. + \sum\limits_{k=2}^nk(k-1)\bar{a}_kx^{k-2}\right] - e^{-x}L  \\
	\frac{p}{m_B} & = e^x\deriv[\epsilon]{x} \\
				  & = e^{(\alpha+1) x}\left[\alpha\sum\limits_{k=0}^n\bar{a}_kx^k + \sum\limits_{k=1}^nk\bar{a}_kx^{k-1}\right] + L \\
	\frac{e}{m_B} & = e^x(\epsilon+1) \\
				  & = e^{(\alpha+1) x}\sum\limits_{k=0}^n\bar{a}_kx^k + e^x(1+d) - L \\
	H & = \ln\left(\frac{e+p}{m_Bn_B}\right) = \ln\left(1+\epsilon+\deriv[\epsilon]{x}\right) \\
	  & = \ln\left(1 + d + {e^{\alpha x}}\left((\alpha+1)\sum\limits_{k=0}^n\bar{a}_kx^k + \sum\limits_{k=1}^nk\bar{a}_kx^{k-1}\right)\right)
\end{align}
\begin{widetext}
\begin{align}
	c_s^2 & = \deriv[p]{e} = \frac{\deriv[\epsilon]{x} + \derivdeux[\epsilon]{x}}{1+\epsilon+\deriv[\epsilon]{x}} \\
		  & = \frac{e^{\alpha x}\left[\alpha(\alpha+1)\sum\limits_{k=0}^n\bar{a}_kx^k + (2\alpha+1)\sum\limits_{k=1}^nk\bar{a}_kx^{k-1} + \sum\limits_{k=2}^nk(k-1)\bar{a}_kx^{k-2}\right]}{1 + {d +} e^{\alpha x}\left[(\alpha+1)\sum\limits_{k=0}^n\bar{a}_kx^k + \sum\limits_{k=1}^nk\bar{a}_kx^k\right]} \\
	\Gamma_1 & = \deriv[\ln{p}]{x} = 1 + \frac{\derivdeux[\epsilon]{x}}{\deriv[\epsilon]{x}} \label{e:def_Gamma1}\\
           & = \frac{e^{\alpha x}\left[\alpha(\alpha+1)\sum\limits_{k=0}^n\bar{a}_kx^k + (2\alpha+1)\sum\limits_{k=1}^nk\bar{a}_kx^{k-1} + \sum\limits_{k=2}^nk(k-1)\bar{a}_kx^{k-2}\right]}{e^{\alpha x}\left[\alpha\sum\limits_{k=0}^n\bar{a}_kx^k + \sum\limits_{k=1}^nk\bar{a}_kx^{k-1}\right] + e^{-x}L}
\end{align}
\end{widetext}

\section{Evolution equation for the electron fraction}
\label{app:evolutionelectronfraction}
{Under the 3+1 decomposition of General Relativity, if $\gamma_{ij}$ denotes the induced 3-metric on spacelike hypersurfaces to which we associate $D_i$ its corresponding covariant derivative,
$N$ the lapse function, $\beta^i$ the shift vector, $U^i$ the Eulerian velocity, $\Gamma=(1-U_iU^i)^{-1/2}$ the Lorentz factor
and we denote by $v^i = NU^i-\beta^i$ the coordinate velocity, the evolution equation of $Y_e$ the electron fraction is:
\begin{equation}
  \partial_tY_e + v^iD_iY_e = \frac{N}{\Gamma}\frac{\sigma}{n_B}.
\end{equation}}
The details on the derivation of this equation can be found in~\citep{servignat_new_2023}.

\section{Fit coefficients}\label{s:fit_coefs}
In this appendix we sum up all the fit coefficients for barotropic EoSs as well as some specific physical quantities.

\begin{table*}[t]
  \centering
  \begin{tabular}{|l|l|l|l|l|l|l|l|l|}
  \hline
  & $\bar{a}_1$ & $\bar{a}_2$ & $\bar{a}_3$ & $\bar{a}_4$ & $\bar{a}_5$ & $\bar{a}_6$ & $\bar{a}_7$ & $\alpha$ \\
  \hline
  \hline
  RG(SLy4)~\citep{gulminelli_unified_2015, chabanat_skyrme_1998, danielewicz_symmetry_2009} & 0.28803 & 0.094928 & 0.0078768 & -0.013071 & 0.0010821 & -0.0010058 & -6.6725$\times 10^{-8}$ & 1.5444 \\
  APR(APR)~\citep{haensel_neutron_2007, akmal_equation_1998} & 0.31909 & 0.12187 & 0.026877 & -0.042652 & 0.016685 & 0.014534 & 0.0044882 & 1.9044\\
  GPPVA(DD2)~\citep{typel_composition_2010, grill_equation_2014,pearson_unified_2018} & 0.49196 & -0.029608 & -0.16060 &  0.090142 &  0.15724 &  0.053312 & 0.0062917 & 1.6799\\
  \hline
  \end{tabular}
  \caption{One-argument EoS fit coefficients for the pseudo-polytrope scheme.}
  \label{tab:coefpseudopolytrope}
  \end{table*}
  
  \begin{table*}[t]
  \centering
  \begin{tabular}{|l|l|l|l|l|}
  \hline
  & $n_{\mathrm{lim},1}\,[\mathrm{fm}^{-3}]$ & $n_{\mathrm{lim},2}\,[\mathrm{fm}^{-3}]$ & $\kappa$ [geom] & $\gamma$ \\
  \hline
  \hline
  RG(SLy4)~\citep{gulminelli_unified_2015, chabanat_skyrme_1998, danielewicz_symmetry_2009} & $10^{-4}$ & $10^{-2}$ & 0.140 & 1.34 \\
  APR(APR)~\citep{haensel_neutron_2007, akmal_equation_1998} & $10^{-4}$ & $10^{-2}$ & 0.178 & 1.34 \\
  GPPVA(DD2)~\citep{typel_composition_2010, grill_equation_2014,pearson_unified_2018} & $10^{-5}$ & $8\times10^{-2}$ & 0.0140 & 1.34 \\
  \hline
  \end{tabular}
  \caption{Fit parameters for one-argument EoSs. $\kappa$ is expressed in the set of geometrized units supplemented with $\Msol=1$.}
  \label{tab:parampseudopolytrope}
  \end{table*}
  
\begin{table*}[t]
\centering
\begin{tabular}{|l|l|l|l|}
\hline
$a_{i}$ & RG(SLy4)~\citep{gulminelli_unified_2015, chabanat_skyrme_1998, danielewicz_symmetry_2009} & APR(APR)~\citep{haensel_neutron_2007, akmal_equation_1998} & GPPVA(DD2)~\citep{typel_composition_2010, grill_equation_2014,pearson_unified_2018} \\
\hline
\hline
$a_{1}$ & 12.28260 & {12.01420} & 11.18370 \\
$a_{2}$ & 13.86452 & {13.84209} & 15.72280 \\
$a_{3}$ & 1.37955 & {1.31315} & 1.02836 \\
$a_{4}$ & 3.50552 & {3.58798} & 3.62161 \\
$a_{5}$ & -30.75774 & {-30.71754} & -28.57321 \\ 
$a_{6}$ & 2.10272 & {2.20118} & 2.08129 \\
$a_{7}$ & 3.70203 &  {3.45345} & 5.16074 \\
$a_{8}$ & 13.87372 & {14.58988} & 14.44345 \\
$a_{9}$ & 30.13271 & {30.15473} & 31.78975 \\
$a_{10}$ & -2.03900 & {-2.03509} & -1.99041 \\
$a_{11}$ & 1.57077 & {1.91274} & 1.62725 \\
$a_{12}$ & 15.28327 & {14.77801} & 13.7790 \\
$a_{13}$ & {0.013287} & {0.023175} & 0.16426 \\
$a_{14}$ & 14.30128 & {14.46573} & 14.27033 \\
$a_{15}$ & 9.076260 & {9.09854} & 3.45961 \\
\hline
\end{tabular}
\caption{One-argument EoS fit coefficients for the Potekhin-Pearson scheme.}
\label{tab:coefpearson}
\end{table*}

\begin{table*}[t]
  \centering
  \begin{tabular}{|l|l|l|l|l|l|l|}
  \hline
  Parametrization & \multicolumn{2}{c|}{RG(SLy4)} & \multicolumn{2}{c|}{APR(APR)} & \multicolumn{2}{c|}{GPPVA(DD2)} \\
  \hline
   & $M_\mathrm{max}$ & $\Lambda_{1.4}$ & $M_\mathrm{max}$ & $\Lambda_{1.4}$ & $M_\mathrm{max}$ & $\Lambda_{1.4}$ \\
  \hline
   Original EoS & 2.050 & 297.0 & 2.188 & 249.1 & 2.418 & 676.8 \\
   Pseudo-polytrope (this work) & 2.050 & 289.5 & 2.188 & 240.2  & 2.418 & 696.4 \\
   Potekhin-Pearson fit (this work) & 2.041 & 289.1 & {2.178} & {240.3} & 2.417 & 683.9 \\
   Piecewise polytropes~\citep{read_constraints_2009} & 2.049 & $\times$ & 2.213 & $\times$ & $\times$ & $\times$\\
   GPP~\citep{oboyle_parametrized_2020} & 2.053 & 310.6 & 2.168* & 255.0 & $\times$ & $\times$\\
   Suleiman et al.~\citep{suleiman_polytropic_2022} & 2.049 & 304.98 & $\times$ & $\times$ & 2.417 & 697.9 \\
  \hline
  \end{tabular}
  \caption{Comparison with existing parametrizations of barotropic EoSs. Data from other parametrizations have been collected from tables available within the articles. 
  $M_\mathrm{max}$ is the maximum mass of the EoS in units of $\Msol$, and $\Lambda_{1.4}$ is the value of the tidal deformability of a $1.4\Msol$ NS. {The value provided in~\citep{oboyle_parametrized_2020}
  for the maximum mass of APR is $2.057\Msol$, but we recomputed it using their coefficients and found the value tabulated here.}}
\label{tab:comparisonparametrizations}
\end{table*}

\end{document}